\documentclass[11pt,a4paper]{article}
\pdfoutput=1 
\usepackage{graphicx}
\usepackage{xcolor}
\usepackage[caption=false]{subfig}
\usepackage{mathrsfs,mathtools}
\usepackage{physics,amssymb}
\usepackage{bm}
\usepackage{braket}
\usepackage{listings}
\usepackage{cases}
\usepackage{comment}
\usepackage{soul}
\usepackage{cancel}
\usepackage{cases}
\usepackage[utf8]{inputenc}
\usepackage{url}
\usepackage{empheq}
\usepackage{longtable}
\usepackage[normalem]{ulem}
\usepackage{xspace}
\usepackage{aas_macros}
\usepackage{setspace}
\usepackage{bigints}
\usepackage{jcappub}

\hypersetup{colorlinks=true
,urlcolor=DARKBLUE
,anchorcolor=DARKBLUE
,citecolor=DARKBLUE
,filecolor=DARKBLUE
,linkcolor=DARKBLUE
,menucolor=DARKBLUE
,linktocpage=true
,pdfproducer=medialab
}

\makeatletter
\gdef\@fpheader{ }
\g@addto@macro\bfseries{\boldmath}
\makeatother

\let\oldsqrt\sqrt
\def\sqrt{\mathpalette\DHLhksqrt}
\def\DHLhksqrt#1#2{%
\setbox0=\hbox{$#1\oldsqrt{#2\,}$}\dimen0=\ht0
\advance\dimen0-0.2\ht0
\setbox2=\hbox{\vrule height\ht0 depth -\dimen0}%
{\box0\lower0.4pt\box2}}

\newcommand{\ee}{e}

\newcommand{\sss}[1]{{\scriptscriptstyle{#1}}}
\newcommand{\boldmathsymbol}[1]{{\ensuremath{\boldsymbol{#1}}}}

\newcommand{\uPl}{\mathrm{Pl}}
\newcommand{\uin}{\mathrm{in}}

\newcommand{\umax}{\mathrm{max}}

\newcommand{\ucl}{\mathrm{cl}}

\newcommand{\usssPl}{\sss{\uPl}}

\newcommand{\bmk}{\boldmathsymbol{k}}
\newcommand{\bmx}{\boldmathsymbol{x}}

\newcommand{\calP}{\mathcal{P}}

\newcommand{\calO}{\mathcal{O}}

\newcommand{\Mpc}{\mathrm{Mpc}}

\newcommand{\Mp}{M_\usssPl}

\newcommand{\beq}{\begin{equation}}
\newcommand{\eeq}{\end{equation}}
\newcommand{\bea}{\begin{equation}\begin{aligned}}
\newcommand{\eea}{\end{aligned}\end{equation}}

\newlength{\wsingfig}
\newlength{\wdblefig}
\newlength{\wquadfig}
\newlength{\wtriplefig}
\newcommand{\Eq}[1]{Eq.~(\ref{#1})}
\newcommand{\Eqs}[1]{Eqs.~(\ref{#1})}
\newcommand{\Fig}[1]{Fig.~{\ref{#1}}}

\newcommand{\Refa}[1]{Ref.~{\cite{#1}}}
\newcommand{\Refs}[1]{Refs.~{\cite{#1}}}
\newcommand{\Sec}[1]{Sec.~\ref{#1}}
\newcommand{\Secs}[1]{Secs.~\ref{#1}}
\newcommand{\App}[1]{Appendix~\ref{#1}}

\newcommand{\ubw}{\mathrm{bw}}

\newcommand{\uwell}{\mathrm{well}}
\newcommand{\uth}{\mathrm{th}}

\newcommand{\Mpl}{M_\mathrm{Pl}}

\newcommand{\PBH}{\mathrm{PBH}}
\newcommand{\DM}{\mathrm{DM}}

\newcommand{\ph}{\mathrm{ph}}
\newcommand{\bw}{\mathrm{bw}}
\newcommand{\FPT}{\mathrm{FPT}}
\newcommand{\FP}{\mathrm{FP}}

\newcommand{\cl}{\mathrm{cl}}
\newcommand{\well}{\mathrm{well}}
\newcommand{\free}{\mathrm{free}}

\newcommand{\RTH}{\mathrm{RTH}}

\newcommand{\IR}{\mathrm{IR}}
\newcommand{\UV}{\mathrm{UV}}
\newcommand{\obs}{\mathrm{obs}}
\newcommand{\efolds}{$e$-folds\xspace}

\newcommand{\ie}{i.e.\@\xspace}
\newcommand{\eg}{e.g.\@\xspace}
\newcommand{\after}{\mathrm{after}}

\newcommand{\calC}{\mathcal{C}}
\newcommand{\calD}{\mathcal{D}}

\newcommand{\frakD}{\mathfrak{D}}
\newcommand{\uf}{\mathrm{f}}

\newcommand{\calL}{\mathcal{L}}

\newcommand{\um}{\mathrm{m}}
\newcommand{\calN}{\mathcal{N}}

\newcommand{\uw}{\mathrm{w}}
\newcommand{\bfx}{{\bm{x}}}
\newcommand{\bfk}{\bm{k}}
\newcommand{\bfy}{\mathbf{y}}
\newcommand{\bmy}{\bm{y}}

\newcommand{\bae}[1]{\begin{align} #1 \end{align}}
\newcommand{\bce}[1]{\begin{cases} #1 \end{cases}}
\newcommand{\beae}[1]{\begin{equation}\begin{aligned} #1 \end{aligned}\end{equation}}

\newcommand{\dps}{\displaystyle}
\newcommand{\bfe}[4]{
\begin{figure} 
	\centering
	\includegraphics[#1]{#2}
	\caption{#3}
	\label{#4}
\end{figure}}

\newcommand{\relmiddle}[1]{\mathrel{}\middle#1\mathrel{}}

\newcommand{\relBigm}[1]{\mathrel{}\Bigm#1\mathrel{}}

\definecolor{MONZA}{HTML}{CF000F}
\definecolor{DARKBLUE}{HTML}{00008b}
\definecolor{DARKMAGENTA}{HTML}{8b008b}

\title{Statistics of coarse-grained cosmological fields in stochastic inflation}
\date{\today}

\author[a,b]{Yuichiro Tada}
\affiliation[a]{Institute for Advanced Research, Nagoya University, \\
Furocho Chikusaku Nagoya, Aichi 464-8601 Japan}
\affiliation[b]{Department of Physics, Nagoya University, \\
Furocho Chikusaku Nagoya, Aichi 464-8602 Japan}
\emailAdd{tada.yuichiro@e.mbox.nagoya-u.ac.jp}
\author[c]{and Vincent Vennin}
\affiliation[c]{Laboratoire Astroparticule et Cosmologie, CNRS Universit\'e de Paris, \\
10 rue Alice Domon et L\'eonie Duquet, 75013 Paris, France}
\emailAdd{vincent.vennin@apc.in2p3.fr}

\abstract{
We present a  generic  framework  to  compute  the  one-point  statistics  of cosmological  perturbations,  when  coarse-grained  at  an  arbitrary  scale $R$,  in  the  presence  of quantum diffusion. Making use of the stochastic-$\delta N$ formalism, we show how it can be related to the statistics of the amount of expansion realised until the scale $R$ crosses out the Hubble radius. This leads us to explicit formulae for the probability density function (PDF) of the curvature perturbation, the comoving density contrast, and the compaction function. We then apply our formalism to the calculation of the mass distribution of primordial black holes produced in a single-field model containing a ``quantum well'' (\ie an exactly flat region in the potential). We confirm that the PDFs feature heavy, exponential tails, with an additional cubic suppression in the case of the curvature perturbation. The large-mass end of the mass distribution is shown to be mostly driven by stochastic-contamination effects, which produce black holes more massive than those naively expected. This work bridges the final gap between the stochastic-inflation formalism and the  calculation  of  the  mass  distribution  of  astrophysical  objects  such  as  primordial  black holes, and opens up various prospects that we finally discuss.
}

\arxivnumber{2111.15280}

\begin{document}

\maketitle
\flushbottom
\section{Introduction}
\label{sec:intro}
Cosmological structures observed in the universe are understood as coming from the gravitational amplification of quantum vacuum fluctuations~\cite{Starobinsky:1979ty, Mukhanov:1981xt, Hawking:1982cz, Starobinsky:1982ee, Guth:1982ec, Bardeen:1983qw} during an era of early accelerated expansion called inflation~\cite{Starobinsky:1980te, Sato:1980yn, Guth:1980zm, Linde:1981mu, Albrecht:1982wi, Linde:1983gd}. Since the primordial amplitude of those fluctuations is constrained to be small in the observed range $[10^{-6}\,\Mpc,10^4\,\Mpc]$ (see for instance \Refa{Emami:2017fiy}), they are often approached with perturbative techniques, where they are described by quantum fields evolving on a fixed homogeneous and isotropic background. Outside the above-mentioned range however, cosmological perturbations may be large, and non-perturbative techniques may be required. At large scales, such techniques would be necessary to describe the structure of the universe beyond the observable horizon, where large deviations from homogeneity and isotropy can take place. At small scales, they are required to describe the large fluctuations that possibly give birth to ultra-compact objects such as Primordial Black Holes (PBHs)~\cite{Hawking:1971ei, Carr:1974nx, Carr:1975qj}. 

One such method is the stochastic-inflation formalism~\cite{Starobinsky:1982ee, Starobinsky:1986fx}. It relies on the separate-universe approach~\cite{Salopek:1990jq, Sasaki:1995aw, Wands:2000dp, PhysRevD.68.103515, Rigopoulos:2003ak, Lyth:2005fi}, according to which, on scales larger than the Hubble radius, the universe can be described by an ensemble of independent, locally homogeneous and isotropic patches. In this setup, within each patch, cosmological perturbations are evolved using standard perturbative techniques. Once their wavelength crosses out the size of the patch (as an effect of the accelerated expansion), they source the patches dynamics through a classical stochastic noise, the statistical properties of which are identified with quantum expectation values of the underlying fields. The universe is thus assumed to be homogeneous and isotropic only locally (\ie at the Hubble-radius scale and below), while its large-scale behaviour can feature wide fluctuations as the difference in the realisations of the stochastic noise accumulates in distant patches. This may be seen as an effective method to non-perturbatively incorporate quantum backreaction in the infra-red sector of the theory. 

In this approach, the statistics of cosmological perturbations on large scales can be reconstructed from the knowledge of the dynamics of the separate-universe patches, and this is the goal of the so-called stochastic-$\delta N$ formalism~\cite{Enqvist:2008kt, Fujita:2013cna, Vennin:2015hra}. It relies on the fact that on super-Hubble scales, the local fluctuation in the amount of expansion between an initial flat spatial hypersurface and a final hypersurface of uniform energy density, is nothing but the curvature perturbation~\cite{Starobinsky:1982ee, Starobinsky:1986fxa, Sasaki:1995aw, Wands:2000dp, Lyth:2004gb}. The elapsed time between those hypersurfaces (measured with the amount of expansion, and more precisely with the number of \efolds $N$) fluctuates since the matter fields evolve according to classical stochastic equations, and its statistics can be obtained using first-passage-time analysis~\cite{Vennin:2015hra, Pattison:2017mbe}. This thus allows one to reconstruct the large-scale statistics of curvature perturbations.

This has been successfully applied to a number of situations, both in the presence of a slow-roll attractor~\cite{Pattison:2017mbe, Biagetti:2018pjj, Ezquiaga:2019ftu} and beyond that simple case~\cite{Firouzjahi:2018vet, Ezquiaga:2018gbw, Pattison:2019hef, Pattison:2021oen, Figueroa:2020jkf, Figueroa:2021zah}. As it stands, the formalism delivers the one-point statistics of the curvature perturbation, $\zeta$, when coarse-grained at the Hubble scale at the end of inflation (or more precisely at the coarse-graining scale of the stochastic formalism, which has to be somewhat larger than the Hubble radius, see \Sec{sec:StochaDeltanN}). In practice, this is however not entirely sufficient, for the two following reasons. First, studying the formation of a given structure of mass $M$ (say a PBH) usually requires to coarse grain the perturbation field over a scale determined by $M$ (which roughly corresponds to the Hubble scale at the time when the Hubble mass equals $M$). It is therefore not enough to know the statistics of $\zeta$ coarse grained at the Hubble scale at the end of inflation, one needs to reconstruct the statistics of $\zeta$ when  coarse-grained at any arbitrary scale. Second, the curvature perturbation is not always the most relevant quantity to discuss the fate of an over-density. In the context of PBHs for instance, it has been argued that the comoving density contrast~\cite{Young:2014ana}, or the compaction function~\cite{Shibata:1999zs, Harada:2015yda, Musco:2018rwt}, are more relevant quantities~\cite{Biagetti:2021eep, Kitajima:2021fpq}. 

The present article proposes to bridge this gap and presents a full stochastic-$\delta N$ derivation of the one-point statistics of the curvature perturbation, the density contrast, and the compaction function, when coarse-grained over an arbitrary scale. Let us note that in this work, we focus on the one-point statistics, and we leave the analysis of multiple-point statistics for future work. A first step in that direction was already taken in \Refa{Ando:2020fjm}, which derived the power spectrum (\ie the second moment of the two-point statistics) in the stochastic-$\delta N$ formalism.

Let us also stress that the analysis of the scale dependence in the classical framework (\ie in the absence of quantum diffusion) is simple since a given scale emerges from the Hubble radius at a given field location along a reference classical trajectory. As a consequence, the properties of cosmological perturbations at a given scale are directly related to local field-space properties (such as the shape of the inflationary potential) at the corresponding field location. In the stochastic picture however, this one-to-one correspondence between physical scale and field- (or phase-)space location is lost, which is what makes the calculation technically challenging. Ultimately, the problem can be solved by properly convolving the relevant distributions against backwards distributions of the field values, which is the main technical task of the present work. 

The rest of the paper is organised as follows. In \Sec{sec:StochaDeltanN}, we review the stochastic-$\delta N$ formalism and the calculation of first-passage time statistics. In \Sec{sec:CG:zeta}, we explain how to compute the one-point distribution of the curvature perturbation, and in \Sec{sec:delta:C}, we extend these considerations to the density contrast and the compaction function. We then apply these results to a toy example in \Sec{sec:QuantumWell}, where the inflationary potential contains a flat region dominated by stochastic effects. This allows us to derive the first consistent prediction for the mass of PBHs in such models, in the presence of quantum diffusion. We summarise our findings in \Sec{sec:conclusion} where we also mention a few future directions. The paper finally ends with \App{app:quantum:well} to which various technical aspects of the calculations presented in \Sec{sec:QuantumWell} are deferred.
\section{The stochastic-$\delta N$ formalism}
\label{sec:StochaDeltanN}
In this section, we introduce the stochastic-$\delta N$ formalism. In order to better highlight its differences with the standard approach, we first recall how the usual treatment of quantum fluctuations in cosmological perturbation theory proceeds. We consider that inflation is driven by one or several scalar fields called ``inflatons'' and organised into the field-space vector $\bm{\phi}=\pqty{\phi^1,\phi^2,\phi^3,\cdots}$. When described in General Relativity, its action reads
\bae{
    S=\int\dd[4]{x}\sqrt{-g}\bqty{\frac{1}{2}\Mp^2R-\frac{1}{2}g^{\mu\nu}G_{IJ}(\bm{\phi})\partial_\mu\phi^I\partial_\nu\phi^J-V(\bm{\phi})},
}
where $\Mp$ is the reduced Planck mass, $g^{\mu\nu}$ and $R$ are the (inverse) spacetime metric and the corresponding Ricci scalar, and $G_{IJ}$ and $V$ are the field-space metric and the scalar potential. Here we allow for an arbitrarily curved field-space manifold to remain as generic as possible. In the standard approach, cosmological perturbations are described by small quantum fluctuations evolving on a spatially homogeneous and isotropic, classical background universe. That is, one splits the inflatons and the metric into a classical and homogeneous part, $\bar{\bm{\phi}}(t)$ and $\bar{g}_{\mu\nu}(t)$, and the quantum perturbations $\delta\hat{\bm{\phi}}(t,\bfx)$ and $\delta\hat{g}_{\mu\nu}(t,\bfx)$, \ie
\bae{
    \hat{\bm{\phi}}(t,\bfx)=\bar{\bm{\phi}}(t)+\delta\hat{\bm{\phi}}(t,\bfx), \qquad\qquad
    \hat{g}_{\mu\nu}(t,\bfx)=\bar{g}_{\mu\nu}(t)+\delta\hat{g}_{\mu\nu}(t,\bfx).
}
The action is then series-expanded as
\bae{
    S[\bm{\phi},g]=S^{(0)}[\bar{\bm{\phi}},\bar{g}]
    +\sum_{n=1}^\infty S^{(n)}_{\bar{\bm{\phi}},\bar{g}}[\delta\hat{\bm{\phi}},\delta \hat{g}],
}
where $S^{(n)}$ gathers all terms of order $n$ in the perturbation fields. 
As the background fields $\bar{\bm{\phi}}$ and $\bar{g}$ are assumed to be classical objects, their dynamics are determined by the Euler--Lagrange equations associated with $S^{(0)}$, which read
\bea
\label{eq:eom:background}
    \bar{\phi}^{I\prime}(N)&=G^{IJ}(\bar{\bm{\phi}})\frac{\bar{\pi}_J}{H} 
    \, ,\qquad\qquad
    D_N\bar{\pi}_I(N)=-3\bar{\pi}_I-\frac{V_I(\bar{\bm{\phi}})}{H}, \\
    3\Mp^2H^2&=\frac{1}{2}G^{IJ}(\bar{\bm{\phi}})\bar{\pi}_I\bar{\pi}_J+V(\bar{\bm{\phi}}).
\eea
Here and hereafter we use the number of \efolds  $N$ as the time variable (so a prime denotes derivation with respect to the number of \efolds), in terms of which the background Friedmann--Lemaitre--Robertson--Walker metric is given by
\bae{
    \dd{s^2}=g_{\mu\nu}\dd{x^\mu}\dd{x^\nu}=-\frac{1}{H^2}\dd{N^2}+a^2(N)\dd{\bfx^2}
    \qquad\text{where}\qquad
    a(N)=a(N=0)\ee^{N}\, ,
}
with $H$ the Hubble parameter. In \Eq{eq:eom:background}, $D$ is the covariant derivative along the curved field space (so $D_N \bar{\pi}_I =  \bar{\pi}_I' - \Gamma_{IJ}^K \bar{\phi}^{J\prime}\bar{\pi}_K$ where $\Gamma_{IJ}^K$ is the Christoffel symbol associated to the field-space metric $G_{IJ}$) and $\bar{\pi}_I$ are the momenta conjugate to $\bar{\phi}_I$.
Substituting the background solution into $S^{(n)}_{\bar{\bm{\phi}},\bar{g}}$, one can (in principle) handle the quantum-field theory of $\delta\hat{\bm{\phi}}$ and $\delta\hat{g}_{\mu\nu}$ at arbitrary order.

The prescription of the above standard approach is perfectly well defined and unambiguous. However, it relies on the strong assumption that the universe is dominated by the homogeneous mode at all scales. As argued in \Sec{sec:intro}, the accelerated expansion amplifies fluctuations in light degrees of freedom beyond the Hubble scales, so there is no generic guarantee that the universe remains homogeneous on distances much larger than the Hubble radius.
An alternative approach is thus to consider that the universe is homogeneous and isotropic only on scales of the order of the Hubble radius, and to split the fields into a classical, coarse-grained part (the infrared  --- IR --- sector) and a quantum, sub-Hubble part (the ultraviolet --- UV --- sector),
\bea
\label{eq:IR:UV:decomp}
    \hat{\bm{\phi}}(N,\bfx)&=\bm{\phi}_\IR(N,\bfx)+\delta\hat{\bm{\phi}}_\UV(N,\bfx)
    ,\qquad 
    \hat{\bm{\pi}}(N,\bfx)=\bm{\pi}^\IR(N,\bfx)+\delta\hat{\bm{\pi}}^\UV(N,\bfx), \\
    \hat{g}_{\mu\nu}(N,\bfx)&=g^\IR_{\mu\nu}(N,\bfx)+\delta \hat{g}^\UV_{\mu\nu}(N,\bfx) .
\eea
In these expressions, the IR part of a generic field $\calO$ is defined by coarse-graining that field over the scale $R=(\sigma H)^{-1}$, where $\sigma$ is a small positive parameter that ensures that the coarse-graining scale is well above the Hubble radius (see \Refa{Grain:2017dqa} for a discussion on the requirements on $\sigma$). In practice, that coarse-graining procedure is performed via a certain window function $W$,
\bea
\label{eq:cg:def}
\calO_R\left(\bfx\right) = \frac{3}{4\pi}\left(\frac{a}{R}\right)^3\int\dd[3]{\bm{y}}\calO(\bm{y})W\left(\frac{a\left\vert \bm{y}-\bm{x}\right\vert}{R}\right)\, ,
\eea 
where $W\simeq1$ for small arguments and $W\simeq0$ otherwise. In other words, $W$ selects out spatial points that are distant from $\bfx$ by less than the distance $R$. The window function should also be normalised in the sense that coarse-graining a uniform field should leave it invariant, which leads to the condition $3\int u^2 W(u) \dd u = 1$. In Fourier space, \Eq{eq:cg:def} can be rewritten as
\bea  
\label{eq:cg:Fourier}
\calO_R\left(\bfx\right) = \int \frac{\dd[3]{k}}{(2\pi)^{3/2}} \calO(\bfk) \ee^{i\bfk\cdot\bfx} \widetilde{W}\left(\frac{kR}{a}\right),
\eea 
where $\calO(\bfk)=(2\pi)^{-3/2}\int\dd[3]{\bfx}\ee^{-i\bfk\cdot\bfx}\calO(\bfx)$ is the Fourier mode of $\calO(\bfx)$ and the Fourier-space window function $\widetilde{W}$ is related to the real-space window function $W$ by
\bea 
\label{eq:tilde:W:def}
\widetilde{W}(z) = \frac{3}{z^3}\int_0^\infty W\left(\frac{u}{z}\right) \sin(u) u \dd{u}\, .
\eea
From this expression and the properties of $W$ one can show that $\widetilde{W}\simeq1$ for small arguments and $\widetilde{W}\simeq0$ otherwise. In other words, $\widetilde{W}$ selects out Fourier modes with wavelengths larger than $R$. The specifics of the coarse-graining procedure will play an important role below, which is why we recalled how it is performed in detail.

Coming back to the decomposition~\eqref{eq:IR:UV:decomp}, where $\mathcal{O}_\IR = \mathcal{O}_{R=(\sigma H)^{-1}}$, it can be substituted back into the (Hamiltonian) action as
\bae{
    S[\bm{\phi},\bm{\pi},g]=S^{(0)}[\bm{\phi}_\IR,\bm{\pi}_\IR,g_\IR]+S^{(\UV)}[\bm{\phi}_\IR,\bm{\pi}_\IR,g_\IR,\delta\hat{\bm{\phi}}_\UV,\delta\hat{\bm{\pi}}_\UV,\delta \hat{g}_\UV] .
}
After integrating out the UV part, one obtains an effective action for the IR sector, which at quadratic order in the UV component leads to the following corrected equations of motion~\cite{Pinol:2020cdp}
\bea
\label{eq: Langevin}
    \frakD_N\phi_\IR^I&=G^{IJ}(\bm{\phi}_\IR)\frac{\pi_J^\IR}{H}+\xi^{QI} 
    \, ,\qquad 
    \frakD_N\pi^\IR_I=-3\pi^\IR_I-\frac{V_I(\bm{\phi}_\IR)}{H}+\xi^P_I, \\
    3\Mp^2H^2&=\frac{1}{2}G^{IJ}(\bm{\phi}_\IR)\pi^\IR_I\pi^\IR_J+V(\bm{\phi}_\IR) .
\eea
In these expressions, $\bm{\xi}^Q$ and $\bm{\xi}^P$ are two (It\^o-type) stochastic Gaussian noises. Being Gaussian, they are fully described by their two-point functions, given by
\bae{\label{eq: noise correlation}
    \braket{\xi^{XI}(N,\bfx)\xi^{YJ}(N^\prime,\bfy)}\simeq A^{XYIJ}\delta(N-N^\prime)\,
    \mathrm{sinc}\left(\sigma aH|\bfx-\bfy|\right)
}
where the noise amplitude is given by the real part of the corresponding power spectrum evaluated at the coarse-graining scale:
\beae{
\label{eq:AXYIJ}
    &A^{XYIJ}=\Re\calP^{XYIJ}(N,k=\sigma aH), \\
    &\braket{X^I(N,\bfk)Y^J(N,\bfk^\prime)}=\frac{2\pi^2}{k^3}\calP^{XYIJ}(N,k)\delta^{(3)}(\bfk+\bfk^\prime).
}
Here, $X$ and $Y$ represent the covariant perturbations $Q$ and $P$, which are related to the fields' UV parts by
\bae{
\label{eq:def:Q:P}
    Q^I\simeq\delta\phi_\UV^I \qc P_I\simeq\delta\pi^\UV_I-\Gamma_{IJ}^K\pi^\IR_K\delta\phi_\UV^J
}
at leading order. Note that in \Eq{eq: noise correlation}, we have assumed that coarse graining is performed via a Heaviside window function in Fourier space, \ie $\widetilde{W}(z)=\theta(1-z)$, which makes the noises white (\ie uncorrelated at different times). The presence of the cardinal sine function in \Eq{eq: noise correlation} also indicates that the realisations of the noises in two distant patches are uncorrelated. For simplicity, one can approximate $\mathrm{sinc}(z)\simeq \theta(1-z)$, hence two spatial points follow the same realisation of the Langevin equations~\eqref{eq: Langevin} as long as their distance is smaller than the coarse-graining scale, and start following independent realisations when they become more distant than that scale. This gives rise to the picture sketched in \Fig{fig:scale}, which we will further comment on below. 
Finally, in \Eq{eq: Langevin}, $\frakD$ denotes the It\^o covariant derivative, so
\beae{
    &\frakD_N\phi^I_\IR=(\phi^I_\IR)'+\frac{1}{2}\Gamma^I_{JK}A^{QQJK}, \\
    &\frakD_N\pi^\IR_I=D_N\pi^\IR_I-\frac{1}{2}\pqty{\Gamma^S_{IJ,K}+\Gamma^M_{IJ}\Gamma^S_{KM}}\pi^\IR_SA^{QQJK}-\Gamma^K_{IJ}A^{QPJ}{}_K\, .
}

As mentioned above, in the stochastic picture, each coarse-grained patch behaves as an independent stochastic process. This is why from the Langevin equations~\eqref{eq: Langevin}, one can derive the equivalent Fokker--Planck equation, which drives the probability density function~(PDF) $P(\bm{\Phi}\mid N)$ associated to the field phase-space coordinates, gathered in the vector $\bm{\Phi}=(\bm{\phi}_\IR,\bm{\pi}_\IR)$ (hereafter we omit the subscript `IR' for brevity), at time $N$. It is given by
\bae{\label{eq: FP}
    \partial_NP(\bm{\Phi}\mid N)&=\calL_\FP(\bm{\Phi})\cdot P(\bm{\Phi}\mid\calN) \nonumber \\
    &=-\calD_{\phi^I}\left({\frac{G^{IJ}}{H}\pi_JP}\right)+\partial_{\pi_I}\bqty{\pqty{3\pi_I+\frac{V_I}{H}}P} \nonumber \\
    &\quad+\frac{1}{2}\calD_{\phi^I}\calD_{\phi^J}(A^{QQIJ}P)+\calD_{\phi^I}\partial_{\pi_J}(A^{QPI}{}_JP)+\frac{1}{2}\partial_{\pi_I}\partial_{\pi_J}(A^{PP}{}_{IJ}P),
}
which defines the Fokker--Planck operator $\calL_\FP$, and where
$\calD$ denotes the phase-space covariant derivative, $\calD_{\phi^I}\calO^{JK\cdots}=D_{\phi^I}\calO^{JK\cdots}+\Gamma_{IR}^S\pi_S\partial_{\pi_R}\calO^{JK\cdots}$.

As explained in \Sec{sec:intro}, a quantity of great interest is the time $\calN$ elapsed between a given fixed point in field phase space and the end of inflation,\footnote{In principle, the final hypersurface should be of uniform energy density in the $\delta N$ formalism, which is not necessarily the case of the end-of-inflation surface. However, since the stochastic noise is turned off at the end of inflation (quantum fluctuations do not cross out the Hubble radius anymore), the number of \efolds that is realised between the end-of-inflation surface and a subsequent hypersurface of uniform energy density is a deterministic quantity, so its contribution can be easily incorporated in the calculation.} since according to the $\delta N$ formalism it is related to the curvature perturbation $\zeta$ on large scales. The amount of expansion $\calN$ varies from one realisation to the other, so it is a stochastic quantity endowed with a PDF $P_\FPT(\calN\mid\bm{\Phi})$, which corresponds to the first-passage time (FPT) distribution between the initial field configuration $\bm{\Phi}$ and the end of inflation. One can show that it is driven by the adjoint Fokker--Planck equation~\cite{Vennin:2015hra,Pattison:2017mbe}
\bae{\label{eq: adjoint FP}
    \partial_\calN P_\FPT(\calN\mid\bm{\Phi})&=\calL_\FP^\dagger(\bm{\Phi})\cdot P_\FPT(\calN\mid\bm{\Phi}) \nonumber \\
    &=\frac{G^{IJ}}{H}\pi_J\calD_{\phi^I}P_\FPT-\pqty{3\pi_I+\frac{V_I}{H}}\partial_{\pi_I}P_\FPT \nonumber \\
    &\quad+\frac{1}{2}A^{QQIJ}\calD_{\phi^I}\calD_{\phi^J}P_\FPT+A^{QPI}{}_J\calD_{\phi^I}\partial_{\pi_J}P_\FPT+\frac{1}{2}A^{PP}{}_{IJ}\partial_{\pi_I}\partial_{\pi_J}P_\FPT,
}
which defines the adjoint Fokker--Planck operator $\calL_\FP^\dagger$, and which needs to be solved with the boundary condition $P_\FPT(\calN\mid\bm{\Phi}\in\partial\Omega)=\delta(\calN)$ on the end-of-inflation surface $\partial\Omega$.\footnote{In some cases, \ie if the potential is not steep enough at large-field value, an additional reflective boundary condition may be required, see \Refs{Assadullahi:2016gkk, Vennin:2016wnk}.} Note that \Eq{eq: adjoint FP} also gives rise to recursive partial differential equations for the $n^{\mathrm{th}}$ moments of the FPT distribution, $\braket{\calN^n(\bm{\Phi})}=\int_0^\infty\calN^nP_\FPT(\calN\mid\bm{\Phi})\dd{\calN}$, namely~\cite{Vennin:2015hra}
\bae{\label{eq: recursive PDE}
    \calL_\FP^\dagger\cdot\braket{\calN^n(\bm{\Phi})}=-n\braket{\calN^{n-1}(\bm{\Phi})} 
\qquad\text{where}\qquad 
    \braket{\calN^0(\bm{\Phi})}=1\, .
}

A generic property of the solutions to \Eq{eq: adjoint FP} is the presence of heavy tails~\cite{Pattison:2017mbe, Ezquiaga:2019ftu, Figueroa:2020jkf, Pattison:2021oen, Atal:2019cdz,Atal:2019erb,Biagetti:2021eep,Kitajima:2021fpq}: at large $\mathcal{N}$, the PDF behaves as $P_\FPT\propto \ee^{-\Lambda_0 \mathcal{N}}$, where $\Lambda_0$ depends on the details of the model under consideration. This leads to a substantial enhancement of the probability to produce large values of $\mathcal{N}$ (hence large curvature perturbations) compared to the Gaussian predictions of the standard leading-order approach, and has therefore important consequences for the abundance of extreme objects such as PBHs~\cite{Bullock:1996at,Garcia-Bellido:1996mdl,Ivanov:1997ia,Yokoyama:1998pt,Clesse:2015wea,Kawasaki:2015ppx,Pattison:2017mbe,Ezquiaga:2018gbw,Biagetti:2018pjj,Panagopoulos:2019ail,Ezquiaga:2019ftu,Figueroa:2020jkf}. Let us stress that these exponential tails cannot be properly described by usual, perturbative parametrisations of non-Gaussian statistics (such as those based on computing the few first moments of the distribution and the non-linearity parameters $f_{\mathrm{NL}}$, $g_{\mathrm{NL}}$, etc.), which can only account for polynomial modulations of Gaussian tails. A non-perturbative approach such as the one presented here is therefore necessary. 

As an example, let us mention the simple toy model where inflation is realised by a single scalar field $\phi$, the potential of which has a flat portion $V=V_0$ between $\phi=0$ and $\phi_\uw$ (this model is further discussed in \Sec{sec:QuantumWell}). Then, starting from an initial field value $\phi$ inside this ``quantum well'', the distribution associated with the first exit time at $\phi=0$ is given by~\cite{Pattison:2017mbe}
\bea
P_\FPT(\calN\mid{\Phi})=\frac{\pi^2}{\mu^2}\sum_{n=0}^\infty \left(2n+1\right) \sin\left[\left(2n+1\right) \frac{\pi}{2} \frac{\phi}{\phi_\uw} \right] \ee^{-\frac{\pi^2}{\mu^2}\left(n+\frac{1}{2}\right)^2\calN}
\eea 
if a reflective boundary is placed at $\phi_\uw$ and where we have defined  $\mu^2=24\pi^2\Mp^2\phi_\uw^2/V_0$. At large $\calN$, the term $n=0$ dominates in the above sum, and one obtains an exponential tail. 
\section{Coarse-grained curvature perturbation}
\label{sec:CG:zeta}
\subsection{Coarse-graining in the stochastic formalism}
\label{sec:coarse:graining}
\bfe{width=0.95\hsize}{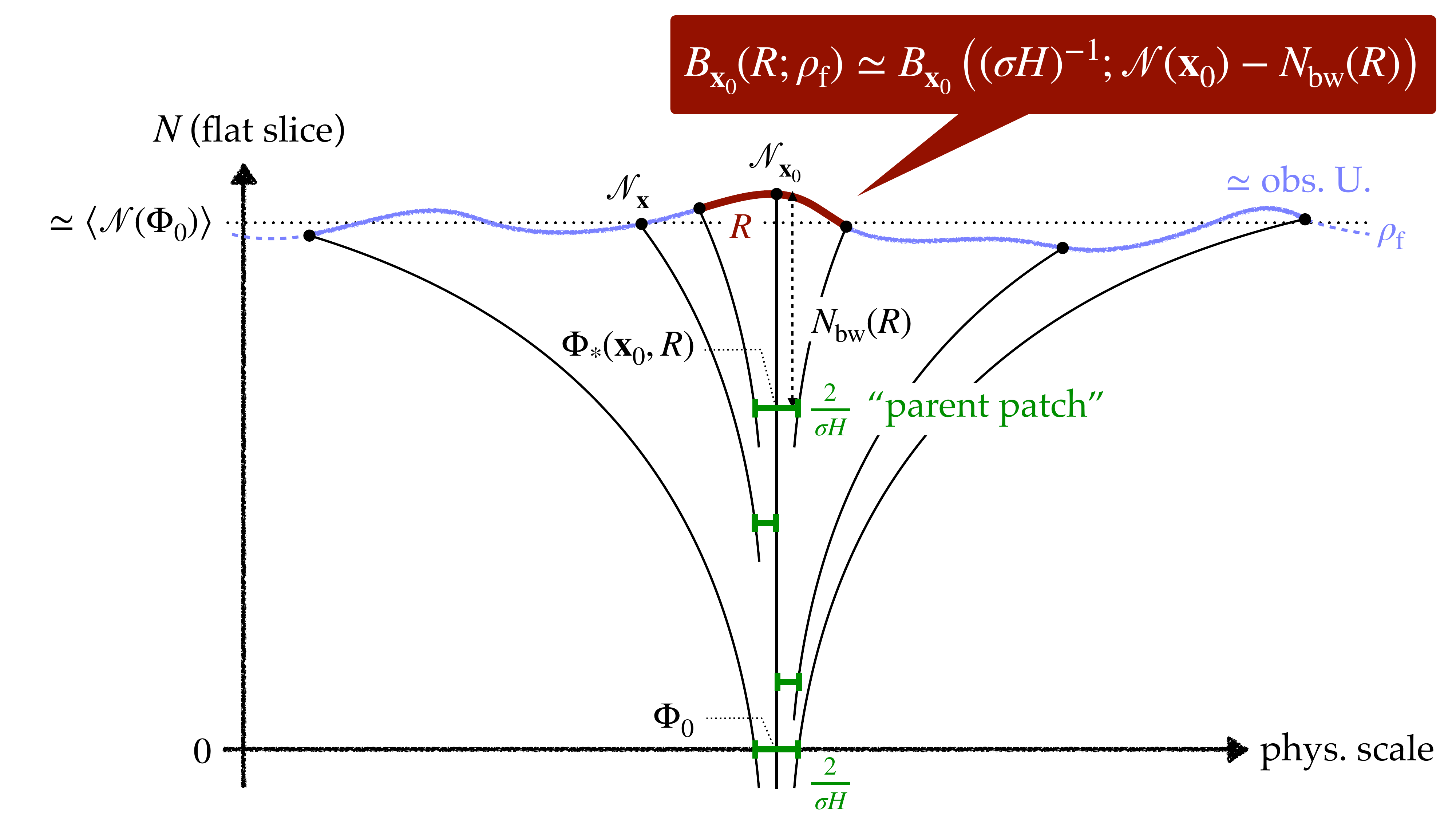}{Space-time diagram sketching the situation considered in this work (see main text).}{fig:scale}
Let us first recall that the stochastic formalism describes the dynamics of quantum fields averaged over the $\sigma$-Hubble scale. As a consequence, the curvature perturbation extracted from the stochastic-$\delta N$ formalism, $\zeta=\calN-\langle\calN\rangle$, is already coarse-grained at the scale $R=(\sigma H_{\mathrm{f}})^{-1}$. Here, $H_{\mathrm{f}}$ corresponds to the Hubble parameter on the final hypersurface of uniform energy density (hence of uniform Hubble parameter, because of Friedmann's equation) on which $\zeta$ is computed. However, as stressed in \Sec{sec:intro}, for practical applications such as the derivation of the abundance of PBHs with a certain mass $M$, one needs to know the one-point statistics of $\zeta$ when coarse-grained at different scales. The calculation needs therefore to be extended to arbitrary $R$, which is the goal of this section.

The situation we consider is depicted in \Fig{fig:scale}. Let $\bfx_0$ label a spatial point on the final hypersurface of constant energy density $\rho=\rho_{\mathrm{f}}$ (displayed with the blue line in \Fig{fig:scale}), on which the curvature perturbation is computed. Let $B_{\bfx_0}(R;\rho_\uf)$ be the set of comoving points $\bfx$ around $\bfx_0$ within the physical distance $R$ on the surface $\rho=\rho_\uf$:
\bae{
\label{eq:zetaR:def}
    B_{\bfx_0}(R;\rho_\uf)=\Bqty{\bfx\mid r_\ph(\bfx,\bfx_0;\rho_\uf)\leq R},
}
where $r_\ph(\bfx,\bfx_0;\rho_\uf)$ denotes the physical distance between $\bfx$ and $\bfx_0$ on the surface $\rho=\rho_\uf$. This set is displayed with the red line in \Fig{fig:scale}, and corresponds to the region over which the curvature perturbation is coarse-grained,\footnote{Let us stress that our formalism can accommodate other averaging procedures. For instance, one may choose to consider the curvature perturbation averaged over \emph{physical} rather than \emph{comoving} coordinates, which implies to insert the scale factor $a^3(\bfx)$ in the integrand of \Eq{eq: volume average of zeta} (as well as in the definition of the volume $V$ given below that, which becomes a physical rather than a comoving volume). Here the scale factor needs to be measured with respect to the observer~\cite{Tada:2016pmk}. Since $a^3(\bfx)=a^3(\bfx_0)\ee^{3(\calN_\bfx-\calN_{\bfx_0})}$, this means that a volume factor should be included in the right-hand side of \Eq{eq:ergodic:appr}, which becomes $\braket{\ee^{3\calN}\calN}$, as well as in what follows. Given that the PDF of $\ee^{3\calN} \calN$ can be straightforwardly obtained from the PDF of $\calN$, one still obtains explicit expressions in terms of first-passage-time statistics, and our formalism can be readily modified to incorporate volume weighting.\label{footnote:VolumeWeighting}}
\bae{\label{eq: volume average of zeta}
    \zeta_R(\bfx_0)=\frac{1}{V[B_{\bfx_0}(R;\rho_\uf)]}\int_{\rho=\rho_\mathrm{f}}\dd[3]{\bfx}\zeta(\bfx)W\left[\frac{r_\ph(\bfx,\bfx_0;\rho_\uf)}{R}\right].
}
Here, $V[B_{\bfx_0}(R;\rho_\uf)]=\int\dd[3]{\bfx} W[r_\ph(\bfx,\bfx_0;\rho_\uf)/R]$ is the comoving volume of $B_{\bfx_0}(R;\rho_\uf)$ and is such that the normalisation condition mentioned below \Eq{eq:cg:def} is satisfied. For notation simplicity, in what follows, the above formula is summarised as  
\bae{
    \zeta_R(\bfx_0)=\braket{\zeta(\bfx)}_{B_{\bfx_0}(R;\rho_\uf)} ,
}
where here and hereafter, the $\langle\cdot\rangle_X$ notation with a subscript $X$ denotes \emph{spatial} average over the set $X$. This should not be confused with the notation $\langle\cdot\rangle$ without subscript, as in \Eq{eq: recursive PDE}, which stands for \emph{stochastic} average.

When $R=R_\obs$, where $R_\obs$ is the size of the observable universe (or more generally the physical size of the region over which observations are performed), $B_{\bfx_0}(R_\obs;\rho_\uf)$ corresponds to the observable universe and is displayed with the blue solid line in \Fig{fig:scale}. As one goes backward in time, the physical size of this (comoving) region decreases, so there is a point at which it matches the $\sigma$-Hubble volume. At this stage, the coarse-grained inflatons are homogeneous across the observable universe and we denote their value by $\bm{\Phi}_0$. This is also where we set the origin of time, $N=0$, and we will see that observable predictions generically depend on $\bm{\Phi}_0$. For each point $\bm{x}$ within the observable universe, one can measure the number of \efolds elapsed between this primeval patch and the final hypersurface, and we denote this quantity by $\calN_{\bfx}(\bm{\Phi}_0)$. According to the $\delta N$ formalism, the curvature perturbation measured within the observable universe on the final hyperspace is given by
\bae{
	\zeta_{(\sigma H_{\mathrm{f}})^{-1}}(\bfx_0)=\calN_{\bfx_0}(\bm{\Phi}_0)-\braket{\calN_\bfx(\bm{\Phi}_0)}_{B_{\bfx_0}(R_\obs;\rho_{\mathrm{f}})}\, .
}
More generally, when the curvature perturbation is coarse-grained at the scale $R$, see \Eq{eq:zetaR:def}, one has
\bea
\label{eq:zetaR:vol:averages}
	\zeta_{R}(\bfx_0)=\braket{\calN_\bfx(\bm{\Phi}_0)}_{B_{\bfx_0}(R;\rho_{\mathrm{f}})}-\braket{\calN_\bfx(\bm{\Phi}_0)}_{B_{\bfx_0}(R_\obs;\rho_{\mathrm{f}})}\, .
\eea 
In principle, this formula is enough to compute the curvature perturbations when coarse-grained at any scale. In a lattice stochastic simulation for instance, it can be simply evaluated if the number of \efolds realised at each node has been properly recorded, by performing ensemble averages over different subsets of the lattice. 
\subsection{From volume averages to stochastic averages}
In order to gain further analytical insight, \Eq{eq:zetaR:vol:averages} has to be cast into a form that allows one to use the techniques presented at the end of \Sec{sec:StochaDeltanN}. The main difficulty is that \Eq{eq:zetaR:vol:averages} is expressed in terms of volume averages, while the first-passage time analysis rather provides stochastic averages. One needs therefore to bridge the gap between those two kinds of averages. 

In order to do so, let us introduce the point at which the physical volume comprised within the set of points $B_{\bfx_0}(R;\rho_{\mathrm{f}})$ matches the $\sigma$-Hubble volume. This corresponds to the upper green segment in \Fig{fig:scale} and below it is referred to as the ``parent patch''. Let $\bm{\Phi}_*(\bfx_0,R)$ denote the value of the inflatons within the parent patch (this value is indeed homogeneous across the patch, since the inflatons are coarse grained at the $\sigma$-Hubble scale). For any point $\bfx$ in $B_{\bfx_0}(R;\rho_{\mathrm{f}})$, one can split the number of \efolds $\calN_{\bfx}(\bm{\Phi}_0)$ between the amount of expansion realised before and after this point,
\bae{
\label{eq:N:split}
	\calN_\bfx(\bm{\Phi}_0)=\calN_\bfx\left[\bm{\Phi}_0\to\bm{\Phi}_*(\bfx_0,R)\right]+\calN_\bfx\left[\bm{\Phi}_*(\bfx_0,R)\right] \qquad \text{for}\qquad \bfx\in B_{\bfx_0}(R;\rho_{\mathrm{f}})\, .
}
Let us look at these two terms separately. Since all points within $B_{\bfx_0}(R;\rho_{\mathrm{f}})$ share the same history prior to the parent patch, the first term is independent of $\bfx$, and one can write
\bea
\left\langle \calN_\bfx\left[\bm{\Phi}_0\to\bm{\Phi}_*(\bfx_0,R)\right] \right\rangle_{B_{\bfx_0}(R;\rho_{\mathrm{f}})} = \calN_{\bfx_0}\left[\bm{\Phi}_0\to\bm{\Phi}_*(\bfx_0,R)\right] .
\eea 
Let us now examine the second term, $\calN_\bfx\left[\bm{\Phi}_*(\bfx_0,R)\right]$. When averaging within $B_{\bfx_0}(R;\rho_{\mathrm{f}})$, one must consider each $\sigma$-Hubble patch comprising the red region in \Fig{fig:scale}, record the number of \efolds that is realised from the parent patch, and take the ensemble average. If all these values were statistically independent, in the limit where there is a large number of such patches, so when $R\gg (\sigma H_{\mathrm{f}})^{-1}$, according to the central-limit theorem the ensemble and stochastic averages would be equal, \ie,
\bea
\label{eq:ergodic:appr}
\left\langle \calN_\bfx\left[\bm{\Phi}_*(\bfx_0,R)\right] \right\rangle_{B_{\bfx_0}(R;\rho_{\mathrm{f}})} \to  \left\langle \calN\left[\bm{\Phi}_*(\bfx_0,R)\right] \right\rangle\, .
\eea 
One may object that in practice, the realisations of $\calN_\bfx\left[\bm{\Phi}_*(\bfx_0,R)\right]$ are not independent for $\bfx\in B_{\bfx_0}(R;\rho_{\mathrm{f}})$, since two comoving points become statistically independent only after their physical distance has grown above $(\sigma H)^{-1}$, see again \Fig{fig:scale}. Those values are therefore correlated. However, the central-limit theorem can be generalised to dependent random variables provided the amount of correlations is bounded (see \eg \Refa{Hoeffding1994}), and it still applies in the present situation. 
Another, maybe more physical, way to understand this result is from the remark that, schematically, the first term in the right-hand side of \Eq{eq:N:split} corresponds to the contributions of scales larger than $R$, while the second term stands for scales smaller than $R$. As explained around \Eq{eq:cg:Fourier}, because of the properties of the Fourier-space window function $\widetilde{W}$, $\zeta_R$ filters out scales smaller than $R$. By performing the replacement~\eqref{eq:ergodic:appr}, one effectively removes the fluctuations in the second term, \ie, one removes the contribution from the scales smaller than $R$. As will be made explicit below, this amounts to using a top-hat window function in Fourier space, which is also required to make the stochastic noises white, see the discussion below \Eq{eq:def:Q:P}. 

This prescription also implies to perform the replacement 
\bea
\braket{\calN_\bfx(\bm{\Phi}_0)}_{B_{\bfx_0}(R_{\mathrm{obs}};\rho_{\mathrm{f}})} \to \braket{\calN(\bm{\Phi}_0) }
\label{eq:ergodic:Robs}
\eea
in \Eq{eq:zetaR:vol:averages}, which corresponds to the mean number of \efolds realised from $\bm{\Phi}_0$. Plugging \Eq{eq:N:split} into \Eq{eq:zetaR:vol:averages}, the above considerations lead to
\bea
\label{eq:zetaR:splitted:clt}
\zeta_R(\bfx_0) = \calN_{\bm{x}_0}\left[\bm{\Phi}_0\to\bm{\Phi}_*(\bfx_0,R)\right] + \left\langle \calN\left[\bm{\Phi}_*(\bfx_0,R)\right] \right\rangle - \braket{\calN(\bm{\Phi}_0)}\, .
\eea 
The only term in the right-hand side of this expression that is subject to stochastic fluctuations is the first one, so schematically, $\zeta_R$ receives contributions from scales comprised between $R$ and $R_\obs$, which is what is expected. Note also that, for a fixed value of $\bm{\Phi}_*$, the problem is cast in terms of first-passage time quantities only,\footnote{One may be concerned that the parent patch does not necessarily corresponds to the first crossing of $\bm{\Phi}_*$. Although this is correct, the time elapsed until crossing the parent patch, $\calN(\bm{\Phi}_0\to \bm{\Phi}_*^{\mathrm{PP}})$ (where ``PP'' stands for ``parent patch''), and the first crossing time of $\bm{\Phi}_*$, $\calN(\bm{\Phi}_0\to \bm{\Phi}_*^{\mathrm{FPT}})$, approximately share the same statistics. This can be seen by  decomposing $\calN(\bm{\Phi}_0)=\calN(\bm{\Phi}_0\to \bm{\Phi}_*^{\mathrm{PP}})+\calN(\bm{\Phi}_*^{\mathrm{PP}}) =\calN(\bm{\Phi}_0\to \bm{\Phi}_*^{\mathrm{FPT}})+\calN(\bm{\Phi}_*^{\mathrm{FPT}})$. Because of the Markovian nature of the process we consider, $\calN(\bm{\Phi}_*^{\mathrm{PP}})$ and $\calN(\bm{\Phi}_*^{\mathrm{FPT}})$ share the same statistics (namely the number of \efolds realised from $\bm{\Phi}_*$ does not depend on whether or not $\bm{\Phi}_*$ has been crossed before). If $\bm{\Phi}_0$ is sufficiently far from $\bm{\Phi}_*$ (so if initial conditions are set at sufficiently early time), they are also uncorrelated with $\calN(\bm{\Phi}_0)$. As a consequence, $\calN(\bm{\Phi}_0\to \bm{\Phi}_*^{\mathrm{PP}})=\calN(\bm{\Phi}_0)-\calN(\bm{\Phi}_*^{\mathrm{PP}})$ and $\calN(\bm{\Phi}_0\to \bm{\Phi}_*^{\mathrm{FPT}})=\calN(\bm{\Phi}_0)-\calN(\bm{\Phi}_*^{\mathrm{FPT}})$ share the same statistics too. \label{footnote:FPT:or:not:FPT:that:is:the:question}} which we know how to compute as explained at the end of \Sec{sec:StochaDeltanN}. The only remaining task is therefore to compute the probability distribution associated with $\bm{\Phi}_*$.
\subsection{Backward probability}
A derivation of the probability distribution associated with $\bm{\Phi}_*$ was presented in \Refa{Ando:2020fjm}, let us recall how it proceeds. Since the size of the parent patch is given by $(\sigma H_*)^{-1}$, and given that each elementary volume within the parent patch expands by an amount controlled by $\ee^{\calN_\bfx}$, the physical volume of $B_{\bfx_0}(R;\rho_{\mathrm{f}})$ reads
\bea 
\label{eq:R3:vol}
    \frac{4\pi}{3}R^3=\frac{\ee^{3N_\bw}}{\sigma^3 H^3(\bm{\Phi}_*)}\int_{B_{\bfx_0}(R;\rho_{\mathrm{f}})} \ee^{3\zeta(\bfx)-3\zeta(\bfx_0)}\dd^3\bfx\, .
\eea 
Here, $N_\bw$ is the number of \efolds elapsed between the parent patch and the final hypersurface along the comoving line labeled by $\bfx_0$. In general, it depends on $\bfx_0$ in a non-trivial way. Two approximations can however be made. First, the integral term in the above expression comes from fluctuations of $\zeta(\bfx)$ within the final patch of size $R$, \ie at scales smaller than $R$. For reasons we have already mentioned, such scales can be discarded, so one can approximate the integral by one. Second, if inflation proceeds in a regime where $H$ is almost a constant (such as in the slow-roll or the ultra-slow roll regimes), then one can approximate $H(\bm{\Phi}_*)\simeq H_{\mathrm{f}}$, which we simply denote $H$. In that limit, $N_\bw$ depends only on $R$, namely 
\bae{
    N_\bw(R)=\ln(\sigma HR).
}
The field configuration $\bm{\Phi}_*$ thus corresponds to the value of the inflatons $N_\bw$ \efolds before the end of inflation. 

As a consequence, the statistics of $\bm{\Phi}_*$ follows the backward (hence the subscript ``bw'') probability distribution,
\bea
P_\bw\left[\bm{\Phi}_* \mid N_{\bw}(R)\right] ,
\eea 
which corresponds to the probability that, $N_\bw$ before the end of inflation, one has $\bm{\Phi}=\bm{\Phi}_*$. As shown in \Refa{Ando:2020fjm}, using Bayes' theorem, it is given by
\bae{\label{eq: Pbw}
    P_\bw(\bm{\Phi}\mid N)=P_\FPT(N\mid\bm{\Phi})\frac{\int_0^\infty\dd{N^\prime}P(\bm{\Phi}\mid N^\prime)}{\int_N^\infty\dd{\calN}P_\FPT(\calN\mid\bm{\Phi}_0)}.
}
In this expression, $P$ is the distribution function associated to the fields value at time $N$ and satisfies the Fokker--Planck equation~\eqref{eq: FP}, while $P_\FPT$ is the first-passage-time distribution that satisfies the adjoint Fokker--Planck equation~\eqref{eq: adjoint FP}. The denominator corresponds to the probability that, starting from $\bm{\Phi}_0$, one realises at least $N$ \efolds (otherwise the backward probability is not defined), and it quickly approaches unity if initial conditions are set sufficiently far from the end-of-inflation surface. 

Having determined how to compute the probability associated to $\bm{\Phi}_*$, let us now come back to \Eq{eq:zetaR:splitted:clt}. It indicates that the probability that $\zeta_R$ falls in the range $[\zeta,\zeta+\dd{\zeta}]$ is the probability that $\calN\left(\bm{\Phi}_0\to\bm{\Phi}_*\right)$ falls in the range $[\zeta-\langle  \calN(\bm{\Phi}_*)\rangle + \langle  \calN(\bm{\Phi}_0)\rangle, \zeta+\dd{\zeta}-\langle  \calN(\bm{\Phi}_*)\rangle + \langle  \calN(\bm{\Phi}_0)\rangle]$, once integrated over $\bm{\Phi}_*$. In other words,
\bea 
\label{eq:Pzeta:zetaR:interm}
	P\left(\zeta_R\right) = \int_{\Omega}\dd{\bm{\Phi}_*} P_{\bw} \Bigl[\bm{\Phi}_* \relBigm{|} N_{\bw}(R)\Bigr] 
	P\Bigl[\calN\left(\bm{\Phi}_0\to\bm{\Phi}_*\right) = \zeta_R-\langle  \calN(\bm{\Phi}_*)\rangle + \langle  \calN(\bm{\Phi}_0)\rangle \relBigm{|} \bm{\Phi}_* \Bigr]
\eea 
where $\Omega$ denotes the inflating domain. Let us further consider the last term in this expression, which is a conditional probability, the condition being that we consider only realisations of the stochastic process that cross $\bm{\Phi}_*$ at least once. In the simple case where the field-phase space is one-dimensional (as in single-field slow-roll inflation), if $\bm{\Phi}_*$ is set between $\bm{\Phi}_0$ and the end-of-inflation surface (which here becomes a single point), all trajectories pass through $\bm{\Phi}_*$. Because of the Markovian nature of the process under consideration, the amount of \efolds elapsed until the first crossing of $\bm{\Phi}_*$ is not correlated with what happens subsequently, hence the condition in this last term can be removed and one simply obtains the first-passage-time probability through $\bm{\Phi}_*$. However, if $\bm{\Phi}_*$ is not set between $\bm{\Phi}_0$ and the end-of-inflation point, or if the field-phase space has more than one dimension, not all stochastic trajectories cross $\bm{\Phi}_*$, and the condition becomes crucial: it indicates that one is dealing with a first-passage time probability, restricted to those trajectories that do cross $\bm{\Phi}_*$. Upon extending the definition of the first-passage time distribution to include that condition,\footnote{In practice, such a first-passage-time distribution can still be computed by solving the adjoint Fokker--Planck equation~\eqref{eq: adjoint FP} with an absorbing condition at $\bm{\Phi}_*$, by adding a ``trapping'' boundary along the end-of-inflation surface (for instance setting the potential to $0$ on that surface, such that the fields cannot escape the trap). The trajectories that do not cross $\bm{\Phi}_*$ end up at the bottom of the trap, hence they do not contribute to finite values of $\calN$ in the first-passage time distribution, which simply needs to be renormalised to account for those missing trajectories.} one finally obtains
\bea 
\label{eq: PzetaR exact}
\boxed{
P\left(\zeta_R\right) = \int_{\Omega}\dd{\bm{\Phi}_*} P_\bw\left[\bm{\Phi}_* \mid N_{\bw}(R)\right] P_{\FPT,\bm{\Phi}_0\to \bm{\Phi}_*}\left[\zeta_R-\langle  \calN(\bm{\Phi}_*)\rangle + \langle  \calN(\bm{\Phi}_0)\rangle  \right]}\, .
\eea 
Since we have explained how to compute the two terms that appear in the right-hand side of this expression, this provides an explicit way to compute the distribution function associated with the curvature perturbation when coarse-grained at an arbitrary scale $R$. This was the goal of this section and it constitutes one of the main results of this paper.
\subsection{Consistency checks}
Before applying the formalism developed above to a concrete example, let us check that previously known results are properly recovered.
\paragraph{Second moment and the power spectrum}
A first consistency check is to verify that the second moment of $\zeta_R$ is consistent with the calculation of the power spectrum presented in \Refa{Ando:2020fjm}. From \Eq{eq: PzetaR exact}, one has
\bea
\label{eq:zeta2:interm}
\left\langle \zeta_R^2\right\rangle =
\int \dd{\zeta_R} P(\zeta_R)\zeta_R^2 =
\int_{\Omega}\dd{\bm{\Phi}_*}P_\bw\left[\bm{\Phi}_* \mid N_{\bw}(R)\right] \left\langle\delta\calN^2 \left(\bm{\Phi}_0\to \bm{\Phi}_*\right) \right\rangle , 
\eea 
where $\delta\calN (\bm{\Phi}_0\to \bm{\Phi}_*)\coloneqq\calN(\bm{\Phi}_0\to\bm{\Phi}_*)+\braket{\calN(\bm{\Phi}_*)}-\braket{\calN(\bm{\Phi}_0)}$.

Let us now reproduce this result by means of the power spectrum. By squaring \Eq{eq:cg:Fourier} and taking the quantum expectation value, one has
\bea 
\label{eq:zeta2:PS}
\left\langle \zeta_R^2\right\rangle = \int \calP_\zeta(k) \widetilde{W}^2\left(\frac{kR}{a}\right)\dd{\ln(k)},
\eea 
where $\calP_\zeta(k) \delta^{(3)}(\bmk+\bmk^\prime) = k^3 \langle \hat{\zeta}_k^2\rangle /(2\pi^2)$ 
is the reduced power spectrum of curvature perturbations. Combining Eqs.~(3.5) and (3.11) of \Refa{Ando:2020fjm}, it is given by
\bea
\label{eq:calP:zeta}
\calP_\zeta(k) = -\int_\Omega\dd\bm{\Phi}_*\left. \frac{\partial P_\bw\left(\bm{\Phi}_* \mid N_{\bw}\right)}{\partial N_\bw}\right\vert_{N_\bw=-\ln(k/k_{\mathrm{f}})} \left\langle\delta\calN^2 \left(\bm{\Phi}_0\to \bm{\Phi}_*\right) \right\rangle,
\eea 
where $k_\mathrm{f}$ denotes the comoving scale that crosses out the $\sigma$-Hubble radius on the final hypersurface. By plugging \Eq{eq:calP:zeta} into \Eq{eq:zeta2:PS}, where a top-hat window function is used in Fourier space,  the integral over $\ln(k)$ can be readily performed and one obtains
\bea
\label{eq:zeta2:interm2}
	\left\langle \zeta_R^2\right\rangle = \int_{\Omega}\dd \bm{\Phi}_* \Bigl\lbrace P_\bw\left[\bm{\Phi}_* \mid N_{\bw}(R)\right]-P_\bw\left(\bm{\Phi}_* \mid \infty\right)\Bigr\rbrace \left\langle\delta\calN^2 \left(\bm{\Phi}_0\to \bm{\Phi}_*\right) \right\rangle  .
\eea 
In this expression, $P_\bw(\bm{\Phi}_* \mid \infty)$ corresponds to the backward probability in the asymptotic past. Since all trajectories originate from $\bm{\Phi}_0$, one has $P_\bw(\bm{\Phi}_* \mid \infty)=\delta(\bm{\Phi}_*-\bm{\Phi}_0)$, hence this term gives a contribution proportional to $\langle\delta\calN^2 (\bm{\Phi}_0\to \bm{\Phi}_0) \rangle = 0 $. This is why \Eqs{eq:zeta2:interm} and~\eqref{eq:zeta2:interm2} coincide, which shows that the two calculations lead indeed to the same result.\footnote{By differentiating \Eq{eq:zeta2:PS} with respect to $R$, one can express the power spectrum in terms of the second moment of the one-point distribution of $\zeta_R$, hence our result can be used to extract the power spectrum and in this sense it is a generalisation of \Refa{Ando:2020fjm}.}
\paragraph{Classical limit}
Let us then consider the classical limit, \ie the regime of low quantum diffusion. In this limit, the backward distribution is nothing but a Dirac distribution centred on the classical path $\bm{\Phi}_\ucl(N)$, \ie
\bea 
P_\bw(\bm{\Phi}\mid N_\bw) = \delta[\bm{\Phi}-\bm{\Phi}_\ucl(N_\mathrm{f}-N_\bw)] .
\eea 
As shown in \Refs{Vennin:2015hra, Assadullahi:2016gkk}, in the classical regime, the first-passage time distribution is a Gaussian,
\bea 
P_{\FPT,\bm{\Phi}_0\to \bm{\Phi}_*}\left[N_\ucl(\bm{\Phi}_0\to \bm{\Phi}_*)+\delta\calN \right] = \frac{\ee^{-\frac{\delta\calN^2}{2\sigma^2_{\bm{\Phi}_0\to \bm{\Phi}_*}}}}{\sqrt{2\pi \sigma^2_{\bm{\Phi}_0\to \bm{\Phi}_*}}} \, ,
\eea 
the width of which is given by the integrated (classical) power spectrum, \ie 
\bea  
\sigma^2_{\bm{\Phi}_0\to \bm{\Phi}_*} = \int_{N_{\mathrm{f}}}^{N_{\bw}(R)}\dd{\ln(k/k_{\mathrm{f}})} \calP_\zeta(k) ,
\eea 
if a top-hat window function is used in Fourier space. One thus recovers the fact that $\zeta_R$ follows a Gaussian distribution centred on zero and with a width given by the quantum expectation value of the second moment of $\hat{\zeta}_R$ computed in the standard approach to cosmological perturbations presented at the beginning of \Sec{sec:StochaDeltanN}.

Let us also note that these two consistency checks confirm that performing the replacement~\eqref{eq:ergodic:appr} amounts to working with a top-hat window function in Fourier space, as mentioned below \Eq{eq:ergodic:appr}.
\section{Coarse-grained density contrast and compaction function}
\label{sec:delta:C}
As explained in \Sec{sec:intro}, the curvature perturbation is not always the best gauge-invariant quantity to discuss the fate of a given over-density. One reason is that, on large scales, where curvature perturbations are conserved, $\zeta$ can be seen as a mere renormalisation of the scale factor as felt by a local observer, and thus it should not affect the collapse dynamics of a local over-density. This is why, in \Refa{Young:2014ana}, it is argued that the comoving density contrast is more relevant, given that according to Poisson equation it is related to the gradient of the curvature perturbation (hence it is less sensitive to large-scale contributions), namely
\bae{
\label{eq:delta:lin:def}
    \delta^{\mathrm{lin}}(\bfx)\simeq-\frac{2(1+w)}{5+3w}\frac{1}{a^2H^2}\nabla^2\zeta(\bfx) .
}
Note that this expression is valid at leading order in cosmological perturbation theory only, hence the superscript ``lin'', and that $w=p/\rho$ denotes the equation-of-state parameter of the background fluid.

If one accounts for the non-perturbative relation between $\zeta$ and $\delta$, one is rather led to the notion of compaction function~\cite{Shibata:1999zs, Harada:2015yda, Musco:2018rwt}. Assuming a spherically-symmetric peak of the curvature perturbation (high peaks are known to be close to spherical symmetry~\cite{1986ApJ...304...15B}), described by the profile $\zeta(r)$ where $r$ is the comoving distance away from the maximum of the peak, the compaction function $\calC(r)$ is defined by the difference between the Misner--Sharp mass contained in the sphere of comoving radius $r$, and the expected mass in the background universe within the same areal radius. It is related to the curvature perturbation via (see, \eg, \Refa{Kitajima:2021fpq}) 
\bae{\label{eq: compaction function}
    \calC(r)=\frac{3(1+w)}{5+3w}\left\lbrace{1-\bqty{1+r\zeta^\prime(r)}^2}\right\rbrace .
}
Contrary to the density contrast, it is conserved on super-Hubble scales. Denoting $r_\um$ the value of $r$ where $\calC$ is maximum, the PBH formation threshold applies to $C(r_\um)$, while the mass of the resultant black hole is related to the one contained within $r_\um$.

Usually, PBH formation criteria take the form of a threshold value (sometimes with critical scaling~\cite{PhysRevLett.70.9}) for the coarse-grained comoving density contrast, $\delta_R$, or as we just mentioned for the maximum compaction function. It is therefore important to derive the one-point statistics of these two quantities, which is the goal of this section.

The main idea is that both quantities collect the fluctuations in the curvature perturbation around the scale of interest (small scales are suppressed by the coarse-graining procedure and large scales do not intervene either, since as argued above they correspond to a local rescaling of the background). As a consequence, we will show that they can be approximated by the difference of the curvature perturbation when coarse grained at two different scales
\bae{
\label{eq:Deltazeta:def}
    \Delta\zeta(\bfx\mid  
    R_1, R_2)\coloneqq\zeta_{
    R_2}(\bfx)-\zeta_{R_1}(\bfx),
}
with $R_2<R<R_1$.
This also picks up fluctuations of $\zeta$ at scales of order $R$ (if $R$, $R_1$ and $R_2$ are of the same order), and we will compute its statistics from the results of \Sec{sec:CG:zeta}.
\subsection{Coarse-shelled curvature perturbation as a proxy }\label{sec: delta as Dzeta}
Since $\Delta\zeta(\bfx\mid R_1,R_2)$ corresponds to the curvature perturbation averaged over the shell region $R_2<a\abs{\bmy-\bmx}<R_1$, in what follows we refer to it as the ``coarse-shelled'' curvature perturbation. In Fourier space, from \Eq{eq:Deltazeta:def} one has $\Delta\zeta(\bfk\mid R_1,R_2) = f_{\Delta\zeta}(\bfk\mid R_1,R_2) \zeta(\bm{k})$, where 
\bea 
\label{eq:coarse:shelled:window}
f_{\Delta\zeta}(\bfk \mid R_1,R_2) = \widetilde{W}\left(\frac{k R_2}{a}\right) -  \widetilde{W}\left(\frac{kR_1}{a}\right)
= \theta\left(\frac{a}{R_2}-k\right)-\theta\left(\frac{a}{R_1}-k\right) .
\eea 
In the second equality, we have used the fact that in the stochastic formalism, a top-hat window function is employed in Fourier space. This shows that $\Delta\zeta$ is made of scales between $R_1$ and $R_2$, as expected.

For the linear density contrast, \Eq{eq:delta:lin:def} leads to $\delta_R^{\mathrm{lin}}(\bfk) = 2(1+w)/(5+3w) f_{\delta}(\bfk\mid R) \zeta(\bm{k}) $ with
\bea 
\label{eq:f:delta}
f_{\delta}(\bfk\mid R) = \left(\frac{kR}{a}\right)^2
 \widetilde{W}_\delta \left(\frac{kR}{a}\right),
 \eea 
where $\widetilde{W}_\delta$ is the window function that is employed to coarse-grain the density contrast. It is a priori different from the window function used to coarse-grain $\zeta$ in the stochastic formalism, and needs to be optimised with respect to the formation criterion of the cosmological structure under consideration (see for instance \Refs{Young:2019osy, Tokeshi:2020tjq}). In practice, one may choose to work with a Gaussian window function, $\widetilde{W}_\delta(z)=\ee^{-z^2/2}$. When $k\ll a/R$, $f_\delta$ is suppressed by the $k^2$ prefactor in \Eq{eq:f:delta} (which is the reason why the comoving density contrast is used rather than the curvature perturbation), while when $k\gg a/R$, $f_\delta$ is suppressed by the (here Gaussian) window function. This shows that $f_\delta$ peaks at the scale $k=a/R$, as announced above.

For the compaction function, one can proceed as follows. Let us first introduce the curvature perturbation coarse-grained with a top-hat function in real space (note that it differs from the curvature perturbation computed in the stochastic-$\delta N$ formalism, which rather uses a top-hat window function in Fourier space)
\bae{
    \zeta_R^\RTH=\frac{3}{R^3}\int_0^R\zeta\pqty{\frac{R^\prime}{a}}{R^\prime}^2\dd{R^\prime} ,
}
where we have assumed that $\zeta$ depends only on the radial comoving coordinate $r=R/a$, and where ``RTH'' stands for ``real top hat''. By differentiating this formula twice with respect to $R$, one obtains
\bae{
    \frac{R}{a}\zeta^\prime\pqty{\frac{R}{a}}=\frac{R^2}{3}\dv[2]{\zeta_R^\RTH}{R}+\frac{4}{3}R\dv{\zeta_R^\RTH}{R}.
}
Let us now Fourier transform this expression with respect to the location of the peak $\bfx$ (away from which the radial coordinate $r$ is defined),
\bae{
    \left[{\frac{R}{a}\zeta^\prime\pqty{\frac{R}{a}}}\right](\bfk)&=\left[\frac{k^2R^2}{3a^2}\widetilde{W}^{\RTH\prime\prime}\pqty{\frac{kR}{a}}+\frac{4}{3}\frac{kR}{a}\widetilde{W}^{\RTH\prime}\pqty{\frac{kR}{a}}\right]\zeta(\bfk) \nonumber \\
    &=-\frac{k^2R^2}{3a^2}\widetilde{W}^\RTH\pqty{\frac{kR}{a}}\zeta(\bfk),
\label{eq:compaction:window:interm}
}
where in the last equation we used the identity $z^2\widetilde{W}^{\RTH\prime\prime}(z)+4z\widetilde{W}^{\RTH\prime}(z)=-z^2\widetilde{W}^\RTH(z)$. This identity can be readily obtained by plugging a top-hat function into the right-hand side of \Eq{eq:tilde:W:def}, which gives $\widetilde{W}^{\RTH}(z)=3(\sin z - z \cos z)/z^3$. The quantity of interest, $r\zeta'(r)$, can therefore be written as $[(R/a) \zeta'(R/a)](\bfk) = -f_{\calC}(\bfk\mid R)\zeta(\bfk)$ with
\bea
f_{\calC}(\bfk\mid R) = \frac{1}{3}
\left(\frac{kR}{a}\right)^2
\widetilde{W}^{\mathrm{RTH}}\left(\frac{kR}{a}\right)\, .
\label{eq:compaction:window:interm:2}
\eea 
When $k\ll a/R$, $f_{\calC}\propto k^2 R^2/a^2$ is suppressed at the same rate as $f_\delta$. However, at small scales when $k\gg a/R$, $f_{\calC}$ is not suppressed. This lack of UV convergence can be traced back to the way the Misner--Sharp mass is defined, which implicitly relies on a real-space top-hat window function. Such window functions are known to produce heavy UV tails, and a natural solution is to define the Misner--Sharp mass with a smoother window function, or, equivalently, to define it from a density field that is already smoothed~\cite{Kalaja:2019uju}.\footnote{Note that such a smoothing procedure of the density field before computing the Misner--Sharp mass should also account for the sub-Hubble evolution of the density contrast, which can be done, at least in the linear theory, by multiplying the Fourier-space window function by the relevant transfer function~\cite{Kalaja:2019uju}.} When doing so, our ability to derive a relation of the form~\eqref{eq:compaction:window:interm} is lost, but at the level of the approximation underlying the present considerations it is enough to simply replace $\widetilde{W}^{\mathrm{RTH}}$ in \Eq{eq:compaction:window:interm:2} by a more generic window function $\widetilde{W}_{\mathcal{C}}$,
\bea
f_{\calC}(\bfk\mid R) \simeq \frac{1}{3}
\left(\frac{kR}{a}\right)^2
\widetilde{W}_{\mathcal{C}}\left(\frac{kR}{a}\right)\, .
\label{eq:compaction:window:interm:3}
\eea 
For simplicity, in what follows, we set $\widetilde{W}_{\mathcal{C}}$ to be a Gaussian function, $\widetilde{W}_{\mathcal{C}}(z)=\ee^{-z^2/2}$, although one should bear in mind that the details of that window function depend on the way the compaction function has been smoothed. With that choice, one has $f_{\calC}(\bfk\mid R) = f_{\delta}(\bfk\mid R)/3$.

Our next task is thus to approximate the effective window function $f_{\delta}(\bfk\mid R)$, given in \Eq{eq:f:delta}, by the one of the coarse-shelled curvature perturbation, see \Eq{eq:coarse:shelled:window}. The reason why this can be done is that, as already mentioned, both these functions select out scales around $k\sim a/R$. This will allow us to assess the one-point statistics of the linear density contrast, and of the compaction function, from the knowledge of the one-point statistics of the coarse-shelled curvature perturbation. The coarse-shelled curvature perturbation has two free parameters, namely $R_1$ and $R_2$. It is convenient to describe them in terms of the two parameters $\alpha$ and $\beta$, defined as $R_2=\alpha R$ and $R_1=\alpha(1+\beta)R$. 
Our goal is thus to approximate
\bea
\label{eq:window:equivalence}
f_{\delta}(\bfk\mid R)  \simeq \gamma f_{\Delta\zeta}\left[\bfk \mid \alpha(1+\beta)R,\alpha R\right],
\eea 
where $\gamma$ is a third free parameter. These three parameters can be set by requiring that both hands of \Eq{eq:window:equivalence} share the same peak location $k_\umax$, as well as the same ``width'' $\sigma^2$ and ``volume'' $V$ of the peak, where
\bae{\label{eq: peak, width, volume}
\sigma^2(f) \coloneqq\int\left[{\ln k-\ln k_{\max}(f)}\right]^2f(k)\dd{\ln k} 
\qquad\text{and}\qquad
V(f) \coloneqq\int f(k)\dd{\ln k}\, .
}
For the left-hand side of \Eq{eq:window:equivalence}, one has $k_\umax(f_\delta) = \sqrt{2}a/R$, $\sigma^2(f_\delta)= (\pi^2+6\gamma_{\mathrm{E}}^2)/24$ where $\gamma_{\mathrm{E}}$ is Euler's constant, and $V(f_\delta)=1$.
For the right-hand side of \Eq{eq:window:equivalence}, one sets $ \ln k_\umax(\gamma f_{\Delta\zeta}) =(\ln[a/(R\alpha)] + \ln\{a/[R\alpha(1+\beta)]\})/2$, which gives rise to $\sigma^2(\gamma f_{\Delta\zeta}) = \gamma\ln^3(1+\beta)/12$ and $V(\gamma f_{\Delta\zeta}) = \gamma\ln(1+\beta)$. By equating the two versions of these three quantities, one obtains
\bea\label{eq: alpha beta gamma}
\alpha &= \frac{\ee^{-\frac{1}{2}\sqrt{\frac{\pi^2+6\gamma_{\mathrm{E}}^2}{2}}}}{\sqrt{2}},\qquad
\beta = \ee^{\sqrt{\frac{\pi^2+6\gamma_{\mathrm{E}}^2}{2}}}-1\, ,\qquad
\gamma =\sqrt{\frac{2}{\pi^2+6\gamma_{\mathrm{E}}^2}}\, .
\eea 
\subsection{Statistics of the coarse-shelled curvature perturbation}
Having determined how the relevant quantities for PBH formation are related to the coarse-shelled curvature perturbation, $\Delta\zeta$, let us now see how the one-point statistics of $\Delta\zeta$ can be extracted from the stochastic-$\delta N$ formalism.  Recalling that $\zeta_R$ is given by \Eq{eq:zetaR:splitted:clt}, from \Eq{eq:Deltazeta:def} one has
\bae{
    \Delta\zeta(\bfx\mid R_1,R_2) 
    &\simeq\mathcal{N}_{\bm{x}}\left[\bm{\Phi}_0 \to \bm{\Phi}_*\pqty{\bfx, R_2}\right]
    -\mathcal{N}_{\bm{x}}\left[\bm{\Phi}_0 \to \bm{\Phi}_*\pqty{\bfx, R_1}\right] \nonumber \\
    &\quad+\braket{\calN\left[\bm{\Phi}_*(\bfx, R_2)\right]}-\braket{\calN\left[\bm{\Phi}_*(\bfx,R_1)\right]} 
   \, .
}
Because of the Markovian nature of the stochastic process we consider, the two first terms can be combined into $\mathcal{N}_\bfx[\bm{\Phi}_*(\bfx,R_1)\to\bm{\Phi}_*(\bfx,R_2) ]$.
Now, given that $\bm{\Phi}_*\pqty{\bfx,R_1}$ is, by definition, the value of the fields at the time $N_{\mathrm{bw}}(R_1) = \ln[\alpha(1+\beta)\sigma HR]$ before the end of inflation, 
while $\bm{\Phi}_*\pqty{\bm{x}, R_2}$ is the value of the fields at the time $N_{\mathrm{bw}}( R_2) = \ln(\alpha\sigma HR)$ before the end of inflation, these first two terms are nothing but $\ln(1+\beta)$ and hence
\bae{\label{eq: Delta zeta}
    \Delta\zeta(\bfx\mid R_1,R_2)\simeq\ln\left(1+\beta\right)
    +\braket{\calN\left[\bm{\Phi}_*(\bfx, R_2)\right]}-\braket{\calN\left[\bm{\Phi}_*(\bfx,R_1)\right]} 
   \, .
}
The two last terms are less straightforward to evaluate, mostly because the field-space positions $\bm{\Phi}_*(\bfx, R_1)$ and $\bm{\Phi}_*(\bfx,R_2)$ are correlated. Therefore, one needs to work out the \emph{joint backward probability} density $P_\bw(\bm{\Phi}_*^{(1)},\bm{\Phi}_*^{(2)}\mid N_\bw^{(1)},N_\bw^{(2)})$, \ie, the probability that the field is at location $\bm{\Phi}_*^{(2)}$ at the time  $N_\bw^{(2)}=\ln(\sigma HR_2)$ before the end of inflation \emph{and} that it is at location $\bm{\Phi}_*^{(1)}$ at the time $N_\bw^{(1)}=\ln(\sigma HR_1)=\ln[(1+\beta)\sigma HR_2]$ before the end of inflation. 
According to Bayes' theorem, it can be expressed as
\bae{
    P_\bw\left(\bm{\Phi}_*^{(1)},\bm{\Phi}_*^{(2)}\relmiddle{|} N_\bw^{(1)},N_\bw^{(2)}\right)=P_\bw\left(\bm{\Phi}_*^{(1)}\relmiddle{|} N_\bw^{(1)}\right)P_\bw\left[\bm{\Phi}_*^{(2)}\relmiddle{|} N_\bw^{(2)},\calN\left(\bm{\Phi}_*^{(1)}\right)=N_\bw^{(1)}\right].
}
The first term is nothing but the single backwards probability that was already calculated in \Eq{eq: Pbw}.
The second term is also a single backwards probability, but with the additional condition that the inflatons were at $\bm{\Phi}_*^{(1)}$ at the time $N_\bw^{(1)}$ before the end of inflation.
In order to relate this probability to quantities we have already computed, let us consider the joint probability that, starting from $\bm{\Phi}_*^{(1)}$ (where we set $N=0$), the first passage time to the end of inflation is $N_\bw^{(1)}$, \emph{and} that the inflatons cross $\bm{\Phi}_*^{(2)}$ at the time $N=\ln(1+\beta)$. This joint probability can be expressed using Bayes' theorem in two different ways, namely
\bae{
    &P\left\lbrace\calN\left(\bm{\Phi}_*^{(1)}\right)=N_\bw^{(1)},\bm{\Phi}\left[N=\ln(1+\beta)\right]=\bm{\Phi}_*^{(2)}\relmiddle{|}\bm{\Phi}(N=0)=\bm{\Phi}_*^{(1)}\right\rbrace \nonumber \\
    &\qquad=P_\bw\left[\bm{\Phi}_*^{(2)}\relmiddle{|}N_\bw^{(2)},\calN\left(\bm{\Phi}_*^{(1)}\right)=N_\bw^{(1)},\bm{\Phi}(N=0)=\bm{\Phi}_*^{(1)}\right] \nonumber 
   \\ &\qquad\quad\times
  P_\FPT\left[\calN\left(\bm{\Phi}_*^{(1)}\right)=N_\bw^{(1)}\relmiddle{|}\bm{\Phi}(N=0)=\bm{\Phi}_*^{(1)}\right] \nonumber \\
    &\qquad=P_\FPT\left\lbrace\calN\left(\bm{\Phi}_*^{(1)}\right)=N_\bw^{(1)}\relmiddle{|}\bm{\Phi}(N=0)=\bm{\Phi}_*^{(1)},\bm{\Phi}\left[N=(1+\beta)\right]=\bm{\Phi}_*^{(2)}\right\rbrace \nonumber \\
    &\qquad\quad\times P\left\lbrace\bm{\Phi}\left[N=\ln(1+\beta)\right]=\bm{\Phi}_*^{(2)}\relmiddle{|}\bm{\Phi}(N=0)=\bm{\Phi}_*^{(1)}\right\rbrace,
}
which leads to the following expression (note that the condition $\calN(\bm{\Phi}_*^{(1)})=N_\bw^{(1)}$ dispenses with the condition $\bm{\Phi}(N=0)=\bm{\Phi}_*^{(1)}$)
\bae{
    &P_\bw\left[\bm{\Phi}_*^{(2)}\relmiddle{|}N_\bw^{(2)},\calN\left(\bm{\Phi}_*^{(1)}\right)=N_\bw^{(1)}\right] \nonumber \\
    &\qquad=P\left\lbrace\bm{\Phi}\left[N=\ln(1+\beta)\right]=\bm{\Phi}_*^{(2)}\relmiddle{|}\bm{\Phi}(N=0)=\bm{\Phi}_*^{(1)}\right\rbrace \nonumber \\
    &\qquad\quad\times\frac{P_\FPT\left\lbrace\calN\left(\bm{\Phi}_*^{(1)}\right)=N_\bw^{(1)}\relmiddle{|}\bm{\Phi}(N=0)=\bm{\Phi}_*^{(1)},\bm{\Phi}\left[N=(1+\beta)\right]=\bm{\Phi}_*^{(2)}\right\rbrace}{P_\FPT\left[\calN\left(\bm{\Phi}_*^{(1)}\right)=N_\bw^{(1)}\relmiddle{|}\bm{\Phi}(N=0)=\bm{\Phi}_*^{(1)}\right]}.
}
The first term in the right-hand side of this expression simply corresponds to the PDF of the inflatons, which is subject to \Eq{eq: FP}, while the denominator corresponds to a first-passage time probability, which is subject to \Eq{eq: adjoint FP}. Only remains the numerator, which, invoking the Markovian nature of the stochastic process under consideration, can be simplified as
\bae{
    &P_\FPT\left\lbrace\calN\left(\bm{\Phi}_*^{(1)}\right)=N_\bw^{(1)}\relmiddle{|}\bm{\Phi}(N=0)=\bm{\Phi}_*^{(1)},\bm{\Phi}\left[N=(1+\beta)\right]=\bm{\Phi}_*^{(2)}\right\rbrace \nonumber \\
    &\qquad=P_\FPT\left\lbrace\calN\left(\bm{\Phi}_*^{(2)}\right)=N_\bw^{(2)}\relmiddle{|}\bm{\Phi}\left[N=\ln(1+\beta)\right]=\bm{\Phi}_*^{(2)}\right\rbrace,
}
which again reduces to a mere first-passage time probability.
Combining these results, one hence obtains the joint backward probability as
\begin{empheq}[box=\fbox]{align}\label{eq: joint Pbw}
    &P_\bw\left(\bm{\Phi}_*^{(1)},\bm{\Phi}_*^{(2)}\relmiddle{|} N_\bw^{(1)},N_\bw^{(2)}\right) \nonumber \\ 
    &=P_\bw\left(\bm{\Phi}_*^{(1)}\relmiddle{|}N_\bw^{(1)}\right)\frac{P_\FPT\left(N_\bw^{(2)}\relmiddle{|}\bm{\Phi}_*^{(2)}\right)}{P_\FPT\left(N_\bw^{(1)}\relmiddle{|}\bm{\Phi}_*^{(1)}\right)}P\left[\bm{\Phi}_*^{(2)}\relmiddle{|}N=\ln(1+\beta),\bm{\Phi}_*^{(1)}\right].
\end{empheq}
This expression is generic and only assumes $N_{\mathrm{bw}}^{(1)}>N_{\mathrm{bw}}^{(2)}$.
According to \Eq{eq: Delta zeta}, the PDF of $\Delta\zeta$ is thus given by
\begin{empheq}[box=\fbox]{align}\label{eq: PDeltaz}
    P(\Delta\zeta)&=\int_\Omega\dd{\bm{\Phi}_*^{(1)}}\dd{\bm{\Phi}_*^{(2)}}
    P_\bw\left(\bm{\Phi}_*^{(1)},\bm{\Phi}_*^{(2)}\relmiddle{|} N_\bw^{(1)},N_\bw^{(2)}\right) \nonumber \\
    &\qquad\times\delta\left[{\Delta\zeta+\Braket{\calN\left(\bm{\Phi}_*^{(1)}\right)}-\Braket{\calN\left(\bm{\Phi}_*^{(2)}\right)}-\ln\left(1+\beta\right)}\right].
\end{empheq}
The two above equations allow one to evaluate the PDF of the coarse-shelled curvature perturbation, hence of the comoving density contrast and of the compaction function as explained in \Sec{sec: delta as Dzeta}. 
Before closing this section, it is worth comparing \Eq{eq: PDeltaz} with the PDF of $\zeta_R$ itself, namely with \Eq{eq: PzetaR exact}. Since $\zeta_R$ involves the stochastic variable $\calN(\bm{\Phi}_0\to\bm{\Phi}_*)$, see \Eq{eq:zetaR:splitted:clt}, its PDF $P(\zeta_R)$ is given by the first-passage-time probability $P_\FPT[\braket{\calN(\bm{\Phi}_0)}-\braket{\calN(\bm{\Phi}_*)}+\zeta_R\mid\bm{\Phi}_0\to\bm{\Phi}_*]$ weighted with the backward probability (\ie~the probability associated with $\bm{\Phi}_*$). In contrast, $\Delta\zeta$ does not explicitly include a stochastic variable, see \Eq{eq: Delta zeta}, but its stochastic nature arises only indirectly through the distribution of the backward fields $\bm{\Phi}_*$'s. Therefore, its PDF is expressed as a Dirac delta distribution, $\delta[{\Delta\zeta+\braket{\calN(\bm{\Phi}_*^{(1)})}-\braket{\calN(\bm{\Phi}_*^{(2)})}-\ln\left(1+\beta\right)}]$, weighted with the (joint) backward probability.
\section{Example: quantum well}
\label{sec:QuantumWell}
\bfe{width=0.7\hsize}{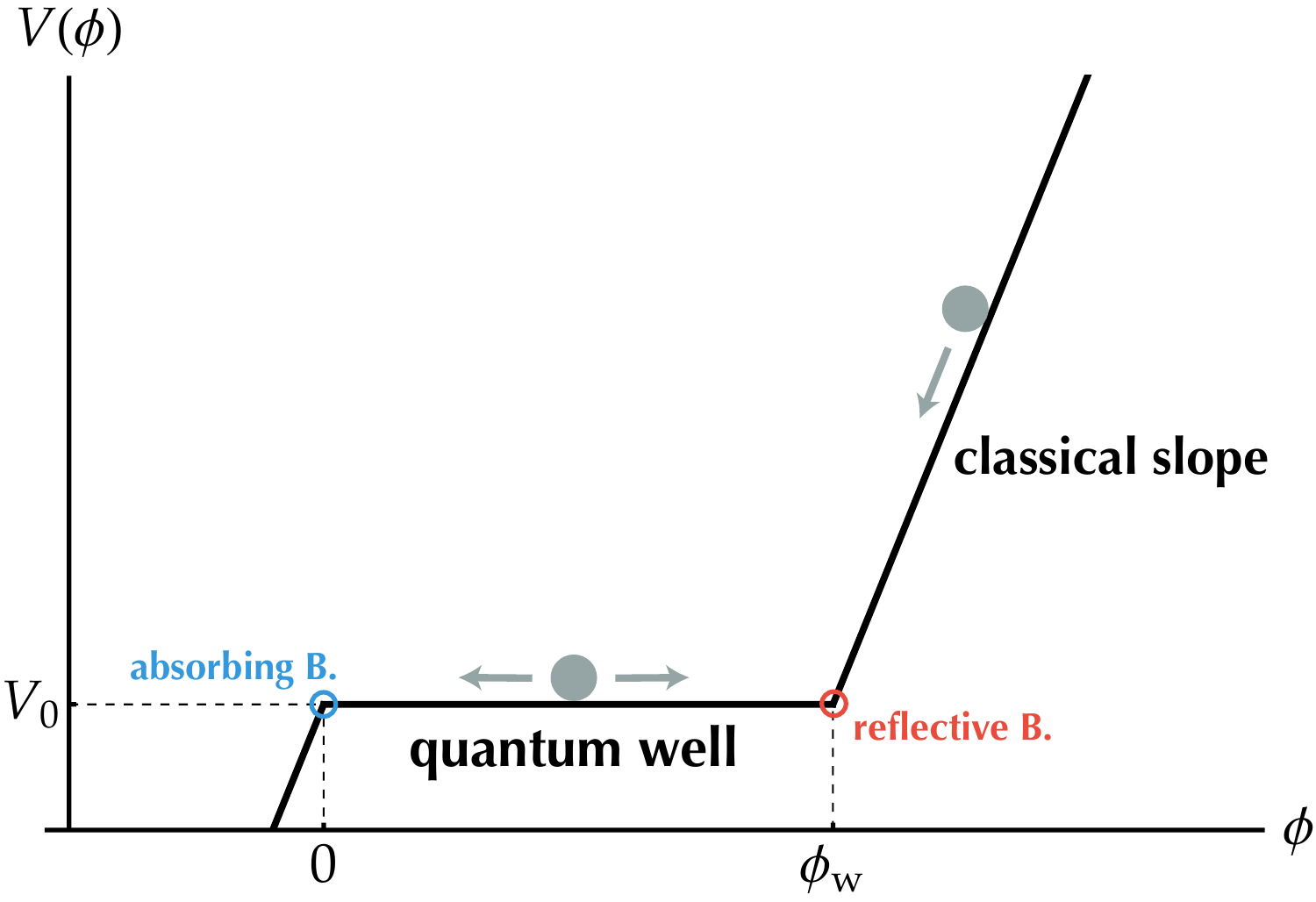}{Schematic representation of the ``quantum well'' toy model studied in \Sec{sec:QuantumWell}, see \Eq{eq: quantum well V}. While the stochastic noise is neglected in the classical region $\phi>\phi_\uw$, the inflaton evolves solely under the action of the quantum noise, \ie~without any potential tilt, in the quantum well $0<\phi<\phi_\uw$. This implies that a reflective boundary is placed at $\phi=\phi_\uw$ once the inflaton is inside the quantum well. Inflation is assumed to end at $\phi=0$, corresponding to an absorbing boundary.}{fig: qwell}
Having established generic formulas to compute the PDF of the coarse-grained curvature perturbation, the density contrast and the compaction function, let us now apply this formalism to a concrete example. This will allow us to illustrate the practical use of our methods, and also to derive a few physical conclusions that should apply more broadly. 

We consider a toy model known as the ``quantum well''~\cite{Pattison:2017mbe,Ezquiaga:2019ftu,Ando:2020fjm} and depicted in \Fig{fig: qwell}. It consists of  a single canonical scalar field $\phi$, whose potential comprises an exactly flat region between $\phi=0$ and $\phi_\uw$ (referred to as the ``quantum well") and a steeper region at $\phi>\phi_\uw$ (the ``classical slope"), 
\bae{\label{eq: quantum well V}
    V(\phi)=\bce{
        V_0, & \text{for $0\leq\phi\leq\phi_\uw$ \quad(quantum well)}, \\
        V_\cl(\phi), & \text{for $\phi>\phi_\uw$ \quad(classical slope)}.
    }
}
In the classical region, the inflaton dynamics is dominated by the potential-induced drift and thus one can neglect the stochastic noise.
Consequently, once the field has landed in the quantum well, the edge $\phi_\uw$ between the classical and quantum regions acts as a reflective boundary (in the language of electric circuits, it is the analogue of a diode).
For the other edge $\phi=0$, which for the time being denotes
the end-of-inflation surface in our setup, we impose an absorbing boundary condition. 

Note that one may add a second classical region below the quantum well~\cite{Ando:2020fjm}, which would shift the number of \efolds~by a constant value. Although it does not affect the amplitude of the curvature perturbation but only shift the scales of perturbations (or the masses of collapsed objects) by a constant value, the effect on the coarse-shelled curvature perturbation $\Delta\zeta$ is more subtle and will be dealt with at the end of \Sec{sec: delta quantum well}.

Assuming that inflation proceeds in the slow-roll regime,\footnote{In principle, the sharp transition between the classical and quantum regions implies that there should be a phase of ultra-slow-roll inflation, the effect of which is further discussed in \Refa{Pattison:2021oen}.} in the quantum well \Eqs{eq: Langevin} and~\eqref{eq:AXYIJ} give rise to $\frakD_N\pi\approx0$, $3\Mp^2H^2\approx V_0$, and $A^{XY}\approx H^2/(2\pi)^2$ for $X=Y=Q$ and $0$ otherwise, while the Langevin equation reads
\bae{
    \dv{x}{N}=\frac{\sqrt{2}}{\mu}\xi(N),\qquad\text{where}\quad
    \braket{\xi(N)\xi(N^\prime)}=\delta(N-N^\prime) .
}   
In this expression, the inflaton has been rescaled according to $x=\phi/\phi_\uw$, and we have introduced the dimensionless parameter $\mu^2=24\pi^2\Mp^2\phi_\uw^2/V_0$.
\subsection{Fokker--Planck and adjoint Fokker--Planck equations}
We first review the analytic solutions of the Fokker--Planck and adjoint Fokker--Planck equations, following \Refs{Pattison:2017mbe,Ando:2020fjm}, since they serve as building blocks of the formulas derived in \Secs{sec:CG:zeta} and~\ref{sec:delta:C}. 

Inside the quantum well, the Fokker--Planck equation~\eqref{eq: FP} reduces to
\bae{\label{eq: FP free}
    \partial_NP(x\mid N)=\frac{1}{\mu^2}\partial_x^2P(x\mid N),
}
which simply describes free Brownian motion. Without any boundary condition, starting from an initial condition $P(x\mid N=0)=\delta(x-x_\uin)$, the solution to \Eq{eq: FP free} is given by 
\bae{\label{eq: Pfree}
    P^\free(x\mid N,x_\uin)=\frac{\mu}{2\sqrt{\pi N}}\ee^{-\frac{\mu^2}{4}\frac{(x-x_\uin)^2}{N}}.
}
The reflective or absorbing conditions can be implemented at the level of the PDF by requiring $\partial_x P=0$ at $\phi=\phi_\uw$ and $P=0$ at $\phi=0$, respectively. The solution satisfying these conditions can be constructed as a linear combination of solutions of the kind~\eqref{eq: FP free}, centred on the various ``images'' of $x_\uin$ across the location of the boundary conditions (this is the so-called ``method of images''), and one obtains~\cite{Ando:2020fjm}
\bae{\label{eq: Pwell}
    P^\well(x\mid N,x_\uin)=\frac{1}{2}\vartheta_2\bqty{-\frac{\pi}{2}(x-x_\uin),\ee^{-\frac{\pi^2N}{\mu^2}}}-\frac{1}{2}\vartheta_2\bqty{-\frac{\pi}{2}(x+x_\uin),\ee^{-\frac{\pi^2N}{\mu^2}}},
}
where $\vartheta_2(z,q)=2\sum_{n=0}^\infty q^{(n+\frac{1}{2})^2}\cos\bqty{(2n+1)z}$ is the second elliptic theta function.

The first-passage-time PDF can be obtained in a similar way by solving the adjoint Fokker--Planck equation~\eqref{eq: adjoint FP}, although it is easier to directly use the above solution of the Fokker--Planck equation and conservation of probability. Indeed, at a given time $N$, the inflaton is either still within the quantum well, or it has been already absorbed at $x=0$, which amounts to
\bae{
    1=\int_0^1P^\well(x\mid N,x_\uin)\dd{x}+\int_0^NP^\well_\FPT(\calN\mid x_\uin,\mu)\dd{\calN}\, .
}
In this expression, $P^\well_\FPT(\calN\mid x_\uin,\mu)$ is the first-passage-time probability from the initial field value $x_\uin$ inside the quantum well, to the end-of-inflation ``surface'' $x=0$.
Differentiating both sides of the above relation with respect to $N$, and making use of the Fokker--Planck equation~\eqref{eq: FP free}, one finds
\bae{
    P^\well_\FPT(\calN\mid x,\mu)=-\frac{1}{\mu^2}\int_0^1\partial_y^2P^\well(y\mid\calN,x)\dd{y}=-\frac{1}{\mu^2}\bqty{\partial_yP^\well(y\mid\calN,x)}_{y=0}^1.
}
The reflective boundary condition at $\phi=\phi_\uw$ ensures that $\eval{\partial_yP^\well(y\mid\calN,x)}_{y=1}=0$, so one obtains
\bae{\label{eq: Pwell FPT}
    P^\well_\FPT(\calN\mid x,\mu)=\frac{1}{\mu^2}\eval{\partial_yP^\well(y\mid\calN,x)}_{y=0}=-\frac{\pi}{2\mu^2}\vartheta^\prime_2\pqty{\frac{\pi}{2}x,\ee^{-\frac{\pi^2\calN}{\mu^2}}}.
}
Here, a prime denotes a derivative with respect to the first argument of the elliptic function. An important property of this PDF is that it features an exponential behaviour in the large $\calN$ limit, 
\bae{\label{eq: exp tail}
    P^\well_\FPT(\calN\mid x,\mu)\underset{\calN\gg\mu^2}{\sim}\frac{\pi}{\mu^2}\sin\pqty{\frac{\pi}{2}x}\ee^{-\frac{\pi^2\calN}{4\mu^2}}.
}
As advocated in \Refa{Ezquiaga:2019ftu} and further checked in \Refs{Figueroa:2020jkf,Pattison:2021oen,Figueroa:2021zah}, these heavy, exponential tails turn out to happen in every model and are not specific to that toy example. This implies that the statistics of the curvature perturbation (and the derived quantities mentioned in \Sec{sec:delta:C}) is endowed with the same heavy, highly non-Gaussian tail behaviour, which has strong implications for the formation of extreme objects such as primordial black holes.

Finally, the classical slope can be incorporated in the analysis by introducing $x_\cl(N;x_\uin)$, which stands for the inflaton solution at time $N$ from $x_\uin>1$ and without stochastic noise; and $N_\cl(x;x_\uin)$, its inverse, that is the number of~\efolds elapsed from $x_\uin$ to $x$ without stochastic noise. When $x_\uin<1$ (that is if one starts inside the quantum well), the solution to the Fokker--Planck equation is still given by \Eq{eq: Pwell}, while for $x_\uin>1$ it reads
\bae{\label{eq: P qwell}
    P(x\mid N,x_0) 
    =\bce{
        \dps
        \delta\left[x-x_\cl(N;x_0)\right] & \qquad \text{if}\quad N<N_\cl(1;x_0) \\
        \dps
        P^\well\left[x\mid N-N_\cl(1;x_0),1\right]& \qquad \text{if}\quad N\geq N_\cl(1;x_0)
    }
    ,
}
where we recall that $P^\well$ is given in \Eq{eq: Pwell}. For the first-passage-time statistics, one has
\bea
\label{eq: Pfpt qwell}
    P_\FPT(\calN\mid x,\mu)=\begin{cases}
    P_\FPT^\uwell(\calN\mid x,\mu) &\qquad \text{if}\quad x<1\\
    P_\FPT^\uwell\left[\calN-N_\cl(1;x)\mid 1,\mu\right]&\qquad \text{if}\quad x\geq 1
\end{cases} ,
\eea 
where we recall that $P_\FPT^\uwell$ is given in \Eq{eq: Pwell FPT}.
\subsection{Coarse-grained curvature perturbation}
The PDF for the coarse-grained curvature perturbation can be obtained by plugging these expressions into \Eqs{eq: Pbw} and \eqref{eq: PzetaR exact}, and in \App{app:quantum:well} we show that it gives rise to
\bae{\label{eq: PzetaR qwell}
    P(\zeta_R)\simeq P^\well(\zeta_R)+\delta(\zeta_R)\int_1^{x_0}\dd{x_*}P_\bw[x_*\mid N_\bw(R)],
}
where $P_\bw$, given in \Eq{eq: Pbw}, is computed in \Eq{eq: Pbw qwell}, and with
\bae{
\label{eq:Pwell:zetaR}
    P^\well(\zeta_R)&=\frac{\pi^2}{4\mu^2}\int_{0}^1 \dd{x_*}\frac{x_*}{(1-x_*)^2}\vartheta_2'\left[{ \frac{\pi}{2}x_*,\ee^{-\frac{\pi^2}{\mu^2}N_\ubw(R)}}\right]\vartheta_2'\bqty{\frac{\pi}{2},e^{-\frac{\pi^2 \zeta_R}{\mu^2(1-x_*)^2}-\frac{\pi^2}{2}}} \nonumber \\
    &\qquad\times \theta\left[{\zeta_R+\frac{\mu^2}{2}(1-x_*)^2}\right].
}
Note that, in the limit where $R=(\sigma H_\mathrm{f})^{-1}$, one recovers the first-passage-time PDF~\eqref{eq: Pwell FPT},\footnote{This can be obtained by first expanding the elliptic function in the limit $a\ll1$ as
\bae{
    x\vartheta_2^\prime\pqty{\frac{\pi}{2}x_*,1-a}\underset{a\ll 1}{\simeq}-x^2\frac{\ee^{-\pqty{\frac{\pi x}{2\sqrt{a}}}^2}}{(a/\pi)^{3/2}}\to -\frac{4}{\pi}\delta(x),
}
where the last expression is derived by integrating both hands against an arbitrary (Taylor-expanded) function, in order to obtain the limit in the space of distributions.
It gives rise to
\bae{
    P^\well(\zeta_R)
    \xrightarrow[N_\bw(R)\to0]{}
    -\frac{\pi^2}{2\mu^2}\vartheta_2^\prime\left[{\frac{\pi}{2},\ee^{-\frac{\pi^2}{\mu^2}\pqty{\zeta_R+\frac{\mu^2}{2}}}}\right],
}
which is nothing but $P^\well_\FPT[\braket{\calN(x=1)}+\zeta_R\mid1,\mu]$, see \Eq{eq: Pwell FPT}.} in agreement with the fact that $\calN-\langle\calN\rangle = \zeta_{(\sigma H_\mathrm{f})^{-1}}$ in stochastic inflation, as pointed out at the beginning of \Sec{sec:coarse:graining}. 
Several comments are in order regarding the above expression.

The first term, $P^\well(\zeta_R)$, corresponds to when the scale $R$ emerges from the Hubble radius when the inflaton is inside the quantum well. The Heaviside function it carries in \Eq{eq:Pwell:zetaR} simply translates the fact that the argument of the first-passage-time PDF appearing in \Eq{eq: PzetaR exact} is positive (namely that the first-passage number of \efolds~is a positive quantity). Let us note that this first term can be written as a universal PDF profile for the rescaled quantity $\zeta_R/\mu^2$, which only depend on one parameter, namely $N_\bw(R)/\mu^2$.

The second term in \Eq{eq: PzetaR qwell} corresponds to when $R$ crosses out the Hubble radius in the classical part of the potential. It is therefore directly proportional to the probability that the inflaton was in the classical part $N_\bw(R)$~\efolds before the end of inflation, to which the integral in \Eq{eq: PzetaR qwell} corresponds. Since we neglected the stochastic noise in the classical slope, it also involves a Dirac distribution $\delta(\zeta_R)$ which forces $\zeta_R$ to vanish. This seemingly divergent part is simply an artefact of this crude assumption, and in practice, accounting for fluctuations in the classical part of the potential would smooth it out. The precise way in which this takes place depends on the details of the potential in the classical region, and since it should only affect the PDF close to its maximum, in what follow we do not further consider that second term when discussing the behaviour of the tail. 

Similarly, it is worth mentioning that the first term in \Eq{eq: PzetaR qwell} also diverges at $\zeta_R=0$, which comes from the rightest edge of the quantum well. This is because $P(\zeta_R)$ involves the first-passage-time PDF $P_{\mathrm{FPT},x_0\to x_*}[\zeta_R-\langle \calN(x_*) \rangle + \langle \calN_{x_*}(x_0)\rangle]$, see \Eq{eq: PzetaR exact}. When $x_*\to 1^-$, the width of the quantum well explored by the path between $x_0$ and $x_*$ decreases to $0$, hence this first-passage-time probability asymptotes $\delta(\zeta_R)$. This can be checked explicitly by evaluating \Eq{eq: Pwell FPT} in the limit $\mu\to 0$. As above, this apparent divergence would be solved by accounting for stochastic fluctuations in the classical slope and a smooth transition with the quantum well. This type of divergence would then only appear from the limiting case $x_*\to x_0$, but this is strongly suppressed by the probability $P_\bw[x_*=x_0\mid N_\bw(R)]$ (or would be smoothed away upon introducing the probability distribution associated to the initial condition $x_0$, since its value remains unknown without further specifying the model).

\bfe{width=0.7\hsize}{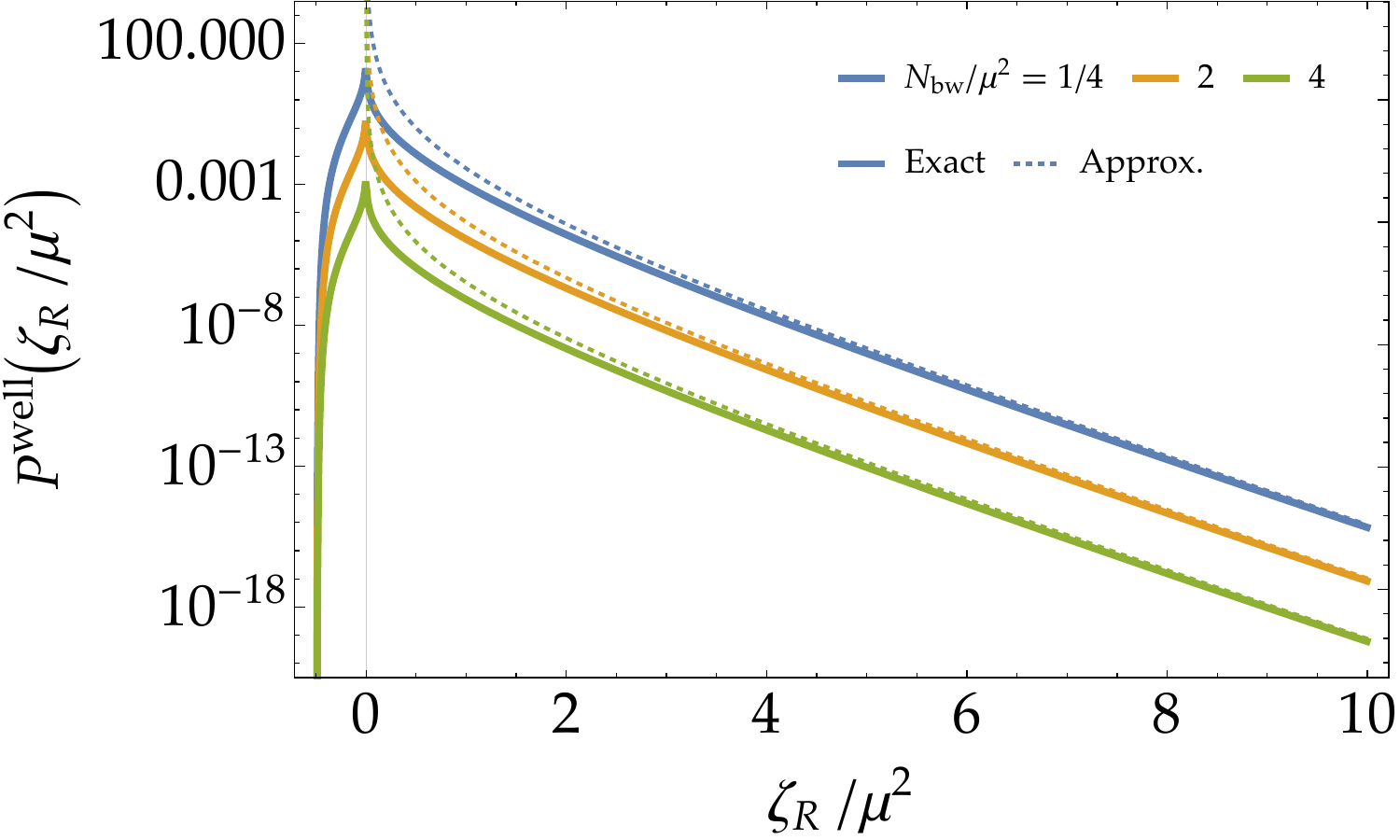}{PDF of the (rescaled) coarse-grained curvature perturbation, $\zeta_R/\mu^2$, as given by the first term in \Eq{eq: PzetaR qwell} (solid lines), and its large-$\zeta_R$ approximation~\eqref{eq: exp tail approx} (dotted lines), for a few values of the coarse-graining scale $R$ labeled with $N_\bw(R)/\mu^2=\ln(\sigma HR)/\mu^2$. One notices the presence of exponential tails at large $\zeta_R$. Note that even though these PDFs do not seem to be normalised, they actually are, due to the presence of an additional Dirac distribution centred at $\zeta_R=0$ that translates the lack of fluctuations in the classical part of the potential, see the discussion in the main text.}{fig: Pwell zR}
The first term of \Eq{eq: PzetaR qwell} is displayed in \Fig{fig: Pwell zR} for a few values of $N_\bw(R)/\mu^2$. One can check that the contribution $P^\well$ from the quantum well remains important even for values of  $N_\bw(R)$ that are larger than the typical number of \efolds~spent in the well, that is $\braket{\calN(x=1)}=\mu^2/2$.
This implies that, even for scales $R$ that emerge in the classical part of the potential with high probability, the far tail is still dominated by those few realisations that do emerge in the quantum well. This is because the heaviness of the tail compensates for the smallness of the probability to emerge in the well. Such a ``contamination" of the non-stochastic part by the quantum well has also been reported in the power spectrum in \Refa{Ando:2020fjm}.

Let us now study the behaviour of the tail of $P(\zeta_R)$. Making use of the asymptotic formula $\vartheta'_2(\pi/2,z) \simeq -2 z^{1/4}$ when $z\ll 1$, which follows from the definition of the $\vartheta_2$ function given below \Eq{eq: Pwell}, in the limit $\zeta_R\gg \mu^2$, the second elliptic function appearing in \Eq{eq:Pwell:zetaR} can be approximated by $-2\ee^{-\pi^2/8}\exp\left[-\frac{\pi^2}{4(1-x_*)^2}\frac{\zeta_R}{\mu^2}\right]$. If $\zeta_R/\mu^2\gg 1$, this function strongly decreases with $x_*$, which is why it can be expanded around $x_*\simeq 0$ where the exponential term can be approximated by $\exp\left[{-\frac{\pi^2}{4}\frac{\zeta_R}{\mu^2}(1+2x_*)}\right]$. Expanding the rest of the integrand in \Eq{eq:Pwell:zetaR} around $x_*=0$, after a couple of integrations by parts (where the boundary terms at $x_*=1$ can be neglected thanks to the exponential suppression in the large $\zeta_R/\mu^2$ limit), one obtains
\bea
\label{eq: exp tail approx}
P\left(\frac{\zeta_R}{\mu^2}\right)\underset{\zeta_R\gg\mu^2}{\sim}
-\frac{4\ee^{-\pi^2/8}}{\pi^3}\vartheta_2^{\prime\prime}\pqty{0,\ee^{-\pi^2\frac{N_\bw(R)}{\mu^2}}}\frac{\ee^{-\frac{\pi^2}{4}\frac{\zeta_R}{\mu^2}}}{\pqty{\zeta_R/\mu^2}^3} .
\eea
This expression is displayed with the dotted lines in \Fig{fig: Pwell zR}, where one can check that it provides a good approximation of the tail. An important remark is that, in addition to the exponential suppression commonly encountered in first-passage-time statistics and already found in \Eq{eq: exp tail}, the tail is further suppressed by the cubic power of $\zeta_R$. This polynomial modulation is a direct effect of the coarse-graining procedure.
\subsection{Coarse-shelled curvature perturbation}\label{sec: delta quantum well}
As explained in \Sec{sec:delta:C}, the one-point statistics of the comoving density contrast and of the compaction function can be inferred from the one of the coarse-shelled curvature perturbation, to which we now turn our attention. It can be obtained by plugging \Eqs{eq: P qwell} and~\eqref{eq: Pfpt qwell} into \Eqs{eq: joint Pbw} and \eqref{eq: PDeltaz}, and in \App{app:quantum:well} we show that the result can be written as
\beae{
\label{eq:P:Deltazeta:quantum:well:gen}
    &P(\Delta\zeta)=P_1(\Delta\zeta)+P_2(\Delta\zeta)+P_3(\Delta\zeta),}
where \beae{
    &P_1(\Delta\zeta)=\frac{\pi}{4\mu^2}\theta\bqty{\ln(1+\beta)-\frac{\mu^2}{2}<\Delta\zeta<\ln(1+\beta)+\frac{\mu^2}{2}} \\
    &\qquad\times\bigintss_{\sqrt{\max\Bqty{0,\frac{2}{\mu^2}\bqty{\Delta\zeta-\ln(1+\beta)}}}}^{\min\Bqty{1,\sqrt{1+\frac{2}{\mu^2}\bqty{\Delta\zeta-\ln(1+\beta)}}}}\dd{\tilde{x}_*^{(1)}}\frac{1-\tilde{x}_*^{(1)}}{\tilde{x}_\well^{(2)}\pqty{\tilde{x}_*^{(1)}}}\vartheta_1^\prime\Bqty{-\frac{\pi}{2}\tilde{x}_\well^{(2)}\pqty{\tilde{x}_*^{(1)}},\ee^{-\frac{\pi^2[N_\bw(R)+\ln\alpha]}{\mu^2}}} \\
    &\qquad\times\pqty{\vartheta_2\Bqty{\frac{\pi}{2}\bqty{\tilde{x}_\well^{(2)}\pqty{\tilde{x}_*^{(1)}}-\tilde{x}_*^{(1)}},\ee^{-\frac{\pi^2\ln(1+\beta)}{\mu^2}}}+\vartheta_2\Bqty{\frac{\pi}{2}\bqty{\tilde{x}_\well^{(2)}\pqty{\tilde{x}_*^{(1)}}+\tilde{x}_*^{(1)}},\ee^{-\frac{\pi^2\ln(1+\beta)}{\mu^2}}}} \\
    &P_2(\Delta\zeta)=\frac{\pi}{2\mu^4}\theta\bqty{-\frac{\mu^2}{2}<\Delta\zeta<\ln(1+\beta)}\bigintss_{\max\bqty{0,\ln(1+\beta)-\Delta\zeta-\frac{\mu^2}{2}}}^{\ln(1+\beta)-\max(0,\Delta\zeta)}\dd{N^{(1)}}\frac{1}{\tilde{x}_\cl^{(2)}\pqty{N^{(1)}}} \\
    &\qquad\times\vartheta_1^\prime\Bqty{-\frac{\pi}{2}\tilde{x}_\cl^{(2)}\pqty{N^{(1)}},\ee^{-\frac{\pi^2\bqty{N_\bw(R)+\ln\alpha}}{\mu^2}}}\vartheta_2\Bqty{\frac{\pi}{2}\tilde{x}_\cl^{(2)}\pqty{N^{(1)}},\ee^{-\frac{\pi^2\bqty{\ln(1+\beta)-N^{(1)}}}{\mu^2}}} \\
    &P_3(\Delta\zeta)=-\frac{\pi}{2\mu^2}\delta(\Delta\zeta)\int_{0}^{N_\bw(R)+\ln\alpha}\dd{N^{(2)}}\vartheta_2^\prime\Bqty{\frac{\pi}{2},\ee^{-\frac{\pi^2\bqty{N_\bw(R)+\ln\alpha-N^{(2)}}}{\mu^2}}},
    \label{eq: P1 P2 P3}
}
with $\tilde{x}_\well^{(2)}=\sqrt{({\tilde{x}_*^{(1)}})^2-\frac{2}{\mu^2}[\Delta\zeta-\ln(1+\beta)]}$ and $\tilde{x}_\cl^{(2)}=\frac{\sqrt{2}}{\mu}\sqrt{\ln(1+\beta)-\Delta\zeta-N^{(1)}}$ and where $\vartheta_1(z,q)=2\sum_{n=0}^\infty (-1)^nq^{(n+\frac{1}{2})^2}\sin\bqty{(2n+1)z}$ is the first elliptic theta function.

The first term, $P_1(\Delta\zeta)$, corresponds to when both $\phi_*^{(1)}$ and $\phi_*^{(2)}$, namely the inflaton values when $R_1$ and $R_2$ respectively cross out the Hubble radius, are in the quantum well; while the second term, $P_2(\Delta\zeta)$, corresponds to when $\phi_*^{(1)}$ is in the classical slope and $\phi_*^{(2)}$ is in the quantum well.
The third term, $P_3(\Delta\zeta)$, which is proportional to the Dirac distribution $\delta(\Delta\zeta_R)$, corresponds to when both $\phi_*^{(1)}$ and $\phi_*^{(2)}$ are in the classical slope. It has the same interpretation as the second term in \Eq{eq: PzetaR qwell} for $P(\zeta_R)$. That is, the Dirac distribution arises from the fact that we have neglected stochastic fluctuations in the classical part of the potential, and it would be smoothed out by accounting for them. The physical interpretation of this term can be made even clearer by noticing the following. Starting from $x_0$, one can always write $\mathcal{N}(x_0) = N_\ucl(x_0)+\mathcal{N}_\uwell$, where $\mathcal{N}_\uwell$ is the number of \efolds spent in the flat well starting from the rightest edge, the PDF of which is given by setting $x=1$ in \Eq{eq: Pwell FPT}, namely $P(\mathcal{N}_\uwell)=P_\FPT^\uwell(\mathcal{N}_\uwell\mid x=1,\mu)$. As a consequence, the probability to be outside the well at the time $N_\ubw^{(2)}$ \efolds before the end of inflation, which we call $1-p_\well(N_\bw^{(2)})$, is given by
\bae{
    1-p_\well\pqty{N_\bw^{(2)}}=P(\mathcal{N}_\uwell<N_\ubw^{(2)}) &= \int_0^{N_\ubw^{(2)}} P_\FPT^\uwell(\mathcal{N}_\uwell\mid x=1,\mu) \dd{\mathcal{N}_\uwell} \nonumber \\
    &=-\frac{\pi}{2\mu^2 }\int_0^{N_\ubw^{(2)}} \vartheta_2'\left(\frac{\pi}{2}x,\ee^{-\frac{\pi^2\mathcal{N}_\uwell}{\mu^2}}\right)\dd{\mathcal{N}_\uwell},
}
where in the last expression we have used \Eq{eq: Pwell FPT}. 
Up to a simple change of variable, $\mathcal{N}_\uwell = N_\bw^{(2)}-N^{(2)}=N_\bw(R)+\ln\alpha-N^{(2)}$, this is precisely what appears in $P_3(\Delta\zeta)$, which can therefore be written as
\bae{
    P_3(\Delta\zeta)=\left[1-p_\uwell\left(N_\bw^{(2)}\right)\right] \delta(\Delta\zeta).
}
If the inflaton is in the classical slope $N_\bw^{(2)}$ \efolds before the end of inflation, it also has to be there $N_\bw^{(1)}$ ($>N_\bw^{(2)}$) \efolds before the end of inflation.  This is why the coarse-shelled curvature perturbation identically vanishes in that case, though this is again simply because we have neglected fluctuations in the classical part. Since resolving this by accounting for the stochastic noise in the classical slope would not affect the tail of the PDF, in what follows we only consider the contribution $P_1(\Delta\zeta)+P_2(\Delta\zeta)$.

\begin{figure}
    \centering
    \begin{tabular}{c}
        \begin{minipage}[b]{0.5\hsize}
            \centering
            \includegraphics[width=0.95\hsize]{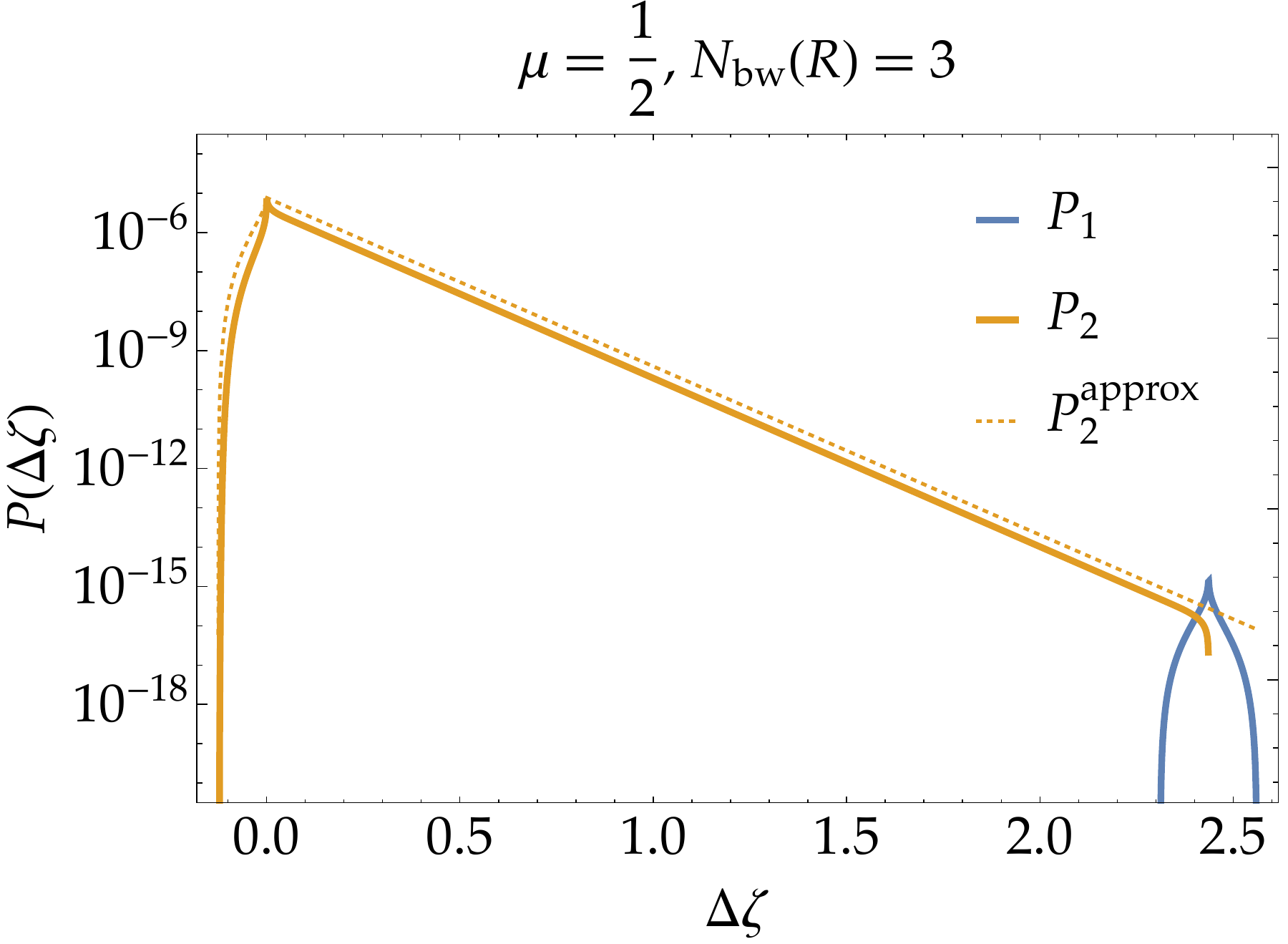}
        \end{minipage}
        \begin{minipage}[b]{0.5\hsize}
            \centering
            \includegraphics[width=0.95\hsize]{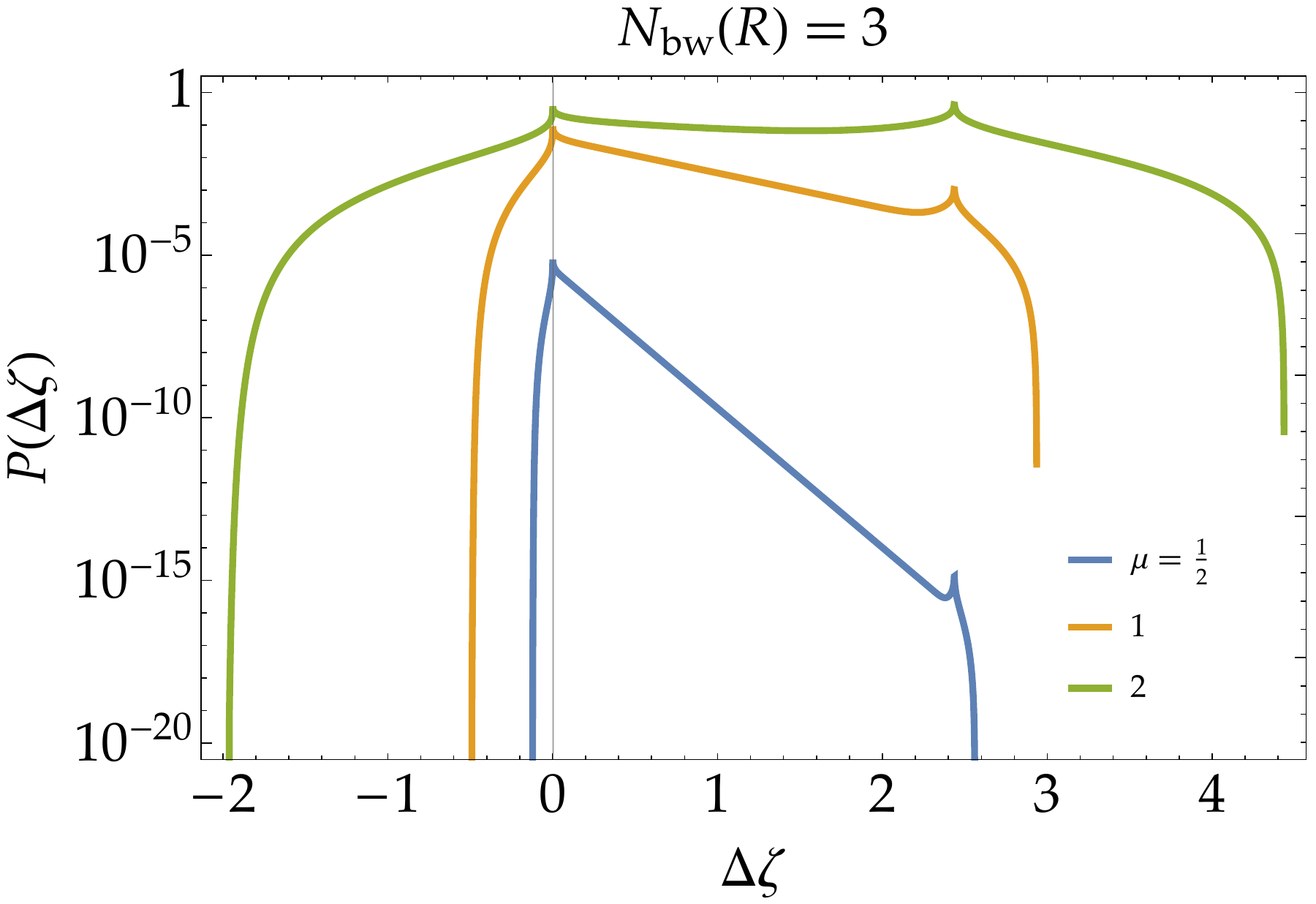}
        \end{minipage}
    \end{tabular}
    \caption{\emph{Left}: The contributions $P_1$ (blue solid line) and $P_2$ (orange solid line) to $P(\Delta\zeta)$ given in \Eq{eq: P1 P2 P3}, for $\mu=1/2$ and $N_\bw(R)=3$, and where the parameters $\alpha$ and $\beta$ are given in \Eq{eq: alpha beta gamma}. The contribution $P_1$ corresponds to when $0<x_*^{(1)},x_*^{(2)}<1$ (\ie both $R_1$ and $R_2$ cross out the Hubble radius in the quantum well), while $P_2$ corresponds to $0<x_*^{(2)}<1<x_*^{(1)}$ (\ie $R_1$ emerges in the classical slope and $R_2$ in the quantum well). One can see that  $P_2$ features an exponential tail, which can be approximated by \Eq{eq:P2:appr} in the small-$\mu$ regime (orange dotted line). It has a hard cutoff at $\Delta\zeta=\ln(1+\beta)$ as an effect of having neglected the stochastic noise in the classical slope. The divergence of $P_1$ at $\Delta\zeta=\ln(1+\beta)$ and of $P_2$ at $\Delta\zeta=0$ are similar artefacts.
    \emph{Right}: Distribution function $P(\Delta\zeta)$ for $N_\bw(R)=3$ and a few values of $\mu$.}
    \label{fig: Nbw3G}
\end{figure}

In the left panel of \Fig{fig: Nbw3G}, we numerically evaluate $P_1$ (dotted line) and $P_2$ (solid line) for $\mu=1/2$ and $N_\bw(R)=3$, where the parameters $\alpha$ and $\beta$ are given in \Eq{eq: alpha beta gamma}. One notices an extended exponential tail supported by $P_2$, which can be approximated by
\bea
\label{eq:P2:appr}
P_2(\Delta\zeta)\simeq \frac{F(1)-F\left[\frac{\sqrt{2\max(0,-\Delta\zeta)}}{\mu}\right]}{\mu^2} \ee^{-\frac{\pi^2}{4\mu^2}\left[N_\bw(R)+\ln(\alpha)+\Delta\zeta\right]}\, ,
\eea
where $F(z)=\sqrt{2\pi}\Bqty{ \mathrm{erf}\qty(\frac{z\pi }{2\sqrt{2}})+\Re\qty[\mathrm{erf}\qty(\frac{2i+z\pi/2}{\sqrt{2}})]\big/e^2}$. This expression is obtained by expanding \Eq{eq: P1 P2 P3} in the regime $\mu\ll 1$, and is displayed with the dotted line in \Fig{fig: Nbw3G}. Already with $\mu=1/2$, it gives a reasonable fit. This exponential tail features a hard cutoff at $\Delta\zeta=\ln(1+\beta)$, around which the contribution $P_1$ gives a substantial correction. This needs to be compared with the PDF of $\zeta_R$, the tail of which is not exactly exponential, see \Eq{eq: exp tail approx}, but which nonetheless undergoes the same exponential suppression.

It is also worth stressing out that $\Delta\zeta$ is bounded from above for both $P_1$ and $P_2$, as explicitly indicated by the step functions in \Eq{eq: P1 P2 P3}. This contrasts with the PDF of $\zeta_R$ itself, see \Eq{eq: PzetaR qwell}, and can be understood as follows. From the definition of $\Delta\zeta$, see \Eq{eq: Delta zeta}, $\Delta\zeta$ is maximum when $x_*^{(2)}$ is maximum and $x_*^{(1)}$ is minimum. For the situations contained in $P_2$, this corresponds to when $x_*^{(1)}=1^+$ and $x_*^{(2)}=1^-$. By continuity of  $\braket{\calN(x)}$, this gives rise to $\Delta\zeta=\ln(1+\beta)$ in \Eq{eq: Delta zeta}, which then acts as a hard cutoff for $P_2$. Similar considerations apply to the cases contained in $P_1$, for which the average number of \efolds are restricted to $0<\braket{\calN(x_*^{(1)})},\braket{\calN(x_*^{(2)})}<\mu^2/2$, and thus $P_1$ takes non-vanishing values only for $\ln(1+\beta)-\mu^2/2<\Delta\zeta<\ln(1+\beta)+\mu^2/2$. This indicates that the compact support of $P(\Delta\zeta)$ is a direct consequence of neglecting the presence of quantum fluctuations in the classical part of the potential, and that the very far tail of $P(\Delta\zeta)$ (\ie above the upper bounds mentioned above) is ultimately driven by those fluctuations. This again contrasts with the tail of $P(\zeta_R)$, which we have shown is driven by the quantum well. 

Another remark of interest is that $P_1$ and $P_2$ have divergent features at $\Delta\zeta=\ln(1+\beta)$ and $\Delta\zeta=0$ respectively. They come from divergences of their integrands at $\tilde{x}_\well^{(2)}=\tilde{x}_\cl^{(2)}=0$,\footnote{The presence of $1/\tilde{x}_\well^{(2)}$ and $1/\tilde{x}_\cl^{(2)}$ in the integrands can be traced back to the Jacobian $[\partial_{x_*^{(2)}}\braket{\calN(x_*^{(2)})}]^{-1}$ coming from the change of variable in the Dirac distribution $\delta[{\Delta\zeta+\braket{\calN(x_*^{(1)})}-\braket{\calN(x_*^{(2)})}-\ln(1+\beta)}]$.} which correspond to the rightest edge of the quantum well $x=1$, and which follow from the reflective boundary condition imposed there, $\eval{\partial_x\braket{\calN(x)}}_{x=1}=0$. Again, by accounting for stochastic diffusion in the classical slope, these features would be smoothed away.

In the right panel of \Fig{fig: Nbw3G}, we show the full PDF $P(\Delta\zeta)$ for a few values of $\mu$.
One can see that $P_1$ can provide the dominant contribution to $P(\Delta\zeta)$ if $\mu^2/2$ is comparable to (or larger than) $N_\bw(R)$. This is because the probability to find both $x_*^{(1)}$ and $x_*^{(2)}$ in the quantum well is substantial in that case.

\bigskip
Before moving on and computing the PBH mass fraction, let us discuss the case where a second classical slope is added below the quantum well, that is at $x<0$. The classical trajectory is denoted $x_\cl^\after(N)$ in that branch. As mentioned above, this simply results in a constant shift in the number of \efolds. It is thus convenient to keep defining the zero point $N_\bw=0$ as corresponding to the lower edge $x=0$, such that scales emerging in the second classical slope, at $x<0$, have $N_\bw<0$. For the coarse-shelled curvature perturbation, the above formulas still apply when $N_\bw^{(1)}$ and $N_\bw^{(2)}$ are positive, but have to be adapted otherwise. Two cases need to be distinguished.

If $N_\bw^{(2)}<N_\bw^{(1)}<0$, both scales emerge in the second classical slope where fluctuations are neglected, hence $\Delta\zeta$ strictly vanishes. More precisely, one has $x_*^{(1)}=x_\cl^\after( -N_\bw^{(1)})$, where $-N_\bw^{(1)}$ is the total number of \efolds realised in the second classical slope, and with a similar expression for $x_*^{(2)}$. One thus has $\langle \calN(x_*^{(1)})\rangle = N_\bw^{(1)}$ and  $\langle \calN(x_*^{(2)})\rangle = N_\bw^{(2)}= N_\bw^{(1)}-\ln(1+\beta)$, so \Eq{eq: Delta zeta} leads to $\Delta\zeta=0$.

If $N_\bw^{(2)}<0<N_\bw^{(1)}$, \ie if only the scale $R_2$ emerges in the second classical slope (but not $R_1$), one still has $x_*^{(2)}=x_\cl^\after( -N_\bw^{(2)})$ and $\langle \calN(x_*^{(2)})\rangle = N_\bw^{(2)}$, and the joint backward probability $P_\bw\left(x_*^{(1)},x_*^{(2)}\relmiddle{|} N_*^{(1)},N_*^{(2)}\right)$ reads
\bae{
    P_\bw\left(x_*^{(1)},x_*^{(2)}\relmiddle{|} N_*^{(1)},N_*^{(2)}\right)=\delta\bqty{x_*^{(2)}-x_\cl^\mathrm{after}\pqty{ -N_\bw^{(2)}}}P_\bw\left(x_*^{(1)}\relmiddle{|}N_\bw^{(1)}\right).
}
In that case, \Eq{eq: Delta zeta} leads to $\Delta\zeta=\ln(1+\beta)+N_\bw^{(2)}-\braket{\calN(x_*^{(1)})}=N_\bw^{(1)}-\braket{\calN(x_*^{(1)})}$, where $\braket{\calN(x_*^{(1)})}$ is given by \Eq{eq: mean N}. If $x_*^{(1)}$ lies in the quantum well, $\braket{\calN(x_*^{(1)})}\leq\mu^2/2$, or equivalently, $\Delta\zeta\geq N_\bw^{(1)}-\mu^2/2$.
The relation between $\Delta\zeta$ and $x_*^{(1)}$ can be inverted as  $x_*^{(1)}=x_\well^{(1)}(\Delta\zeta)\coloneqq1-\sqrt{1-\frac{2}{\mu^2}(N_\bw^{(1)}-\Delta\zeta)}$.
Conversely, if $\Delta\zeta< N_\bw^{(1)}-\mu^2/2$, $x_*^{(1)}$ lies in the first classical slope, and the number of \efolds spent in the quantum well is given by  $N_\bw^{(1)}-\braket{\calN(x_*^{(1)})}+\mu^2/2=\Delta\zeta+\mu^2/2$. Together with the backward probability~\eqref{eq: Pbw qwell}, these considerations lead to
\bae{
    &P(\Delta\zeta)=\int_0^{x_0}\dd{x_*^{(1)}}P_\bw\left(x_*^{(1)}\relmiddle{|}N_\bw^{(1)}\right)\delta\bqty{\Delta\zeta+\braket{\calN(x_*^{(1)})}-N_\bw^{(1)}} \nonumber \\
    &\!=\!\bce{
        \dps
        \!\tilde{P}_1(\Delta\zeta)\!\coloneqq\!-\frac{\pi x_\well^{(1)}(\Delta\zeta)}{2\mu^2\bqty{1-x_\well^{(1)}(\Delta\zeta)}}\vartheta_2^\prime\bqty{\frac{\pi}{2}x_\well^{(1)}(\Delta\zeta),\ee^{-\frac{\pi^2N_\bw^{(1)}}{\mu^2}}} &\text{if } \, N_\bw^{(1)}-\frac{\mu^2}{2}\leq\Delta\zeta<N_\bw^{(1)}, \\
        \dps
        \!\tilde{P}_2(\Delta\zeta)\!\coloneqq\!-\frac{\pi}{2\mu^2}\vartheta_2^\prime\bqty{\frac{\pi}{2},\ee^{-\frac{\pi^2\qty(\Delta\zeta+\frac{\mu^2}{2})}{\mu^2}}} &\text{if } \, -\frac{\mu^2}{2}<\Delta\zeta<N_\bw^{(1)}-\frac{\mu^2}{2}, \\
        0 &\text{otherwise} .
    }
}
Note that, since $0<\braket{\calN(x_*^{(1)})}<N_\bw^{(1)}+\mu^2/2$, the PDF $P(\Delta\zeta)$ is non-zero only for $-\mu^2/2<\Delta\zeta<N_\bw^{(1)}$ if $N_\bw^{(1)}>0$.
\begin{figure}
    \centering
    \begin{tabular}{c}
        \begin{minipage}{0.5\hsize}
            \centering
            \includegraphics[width=0.95\hsize]{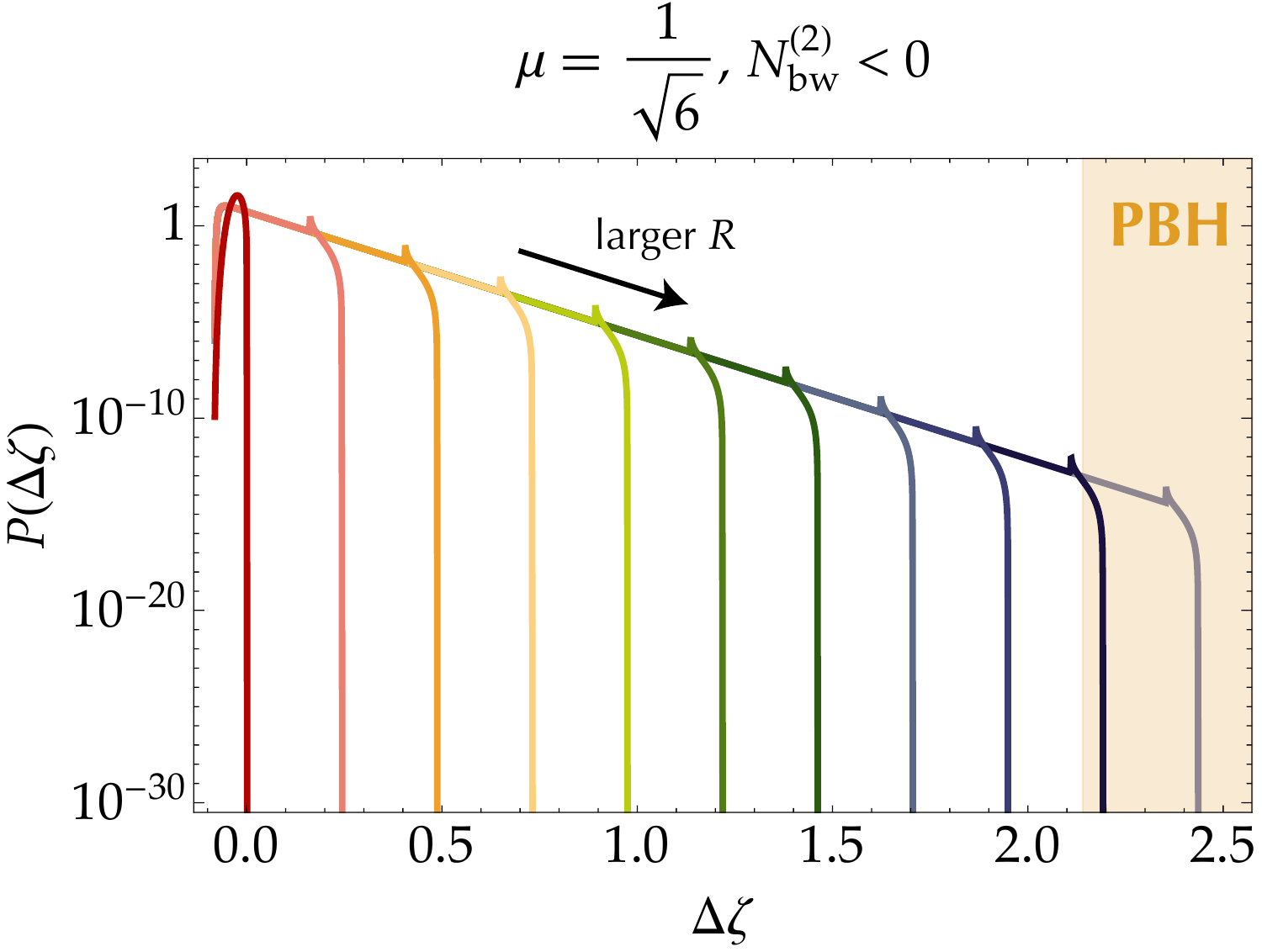}
        \end{minipage}
        \begin{minipage}{0.5\hsize}
            \centering
            \includegraphics[width=0.95\hsize]{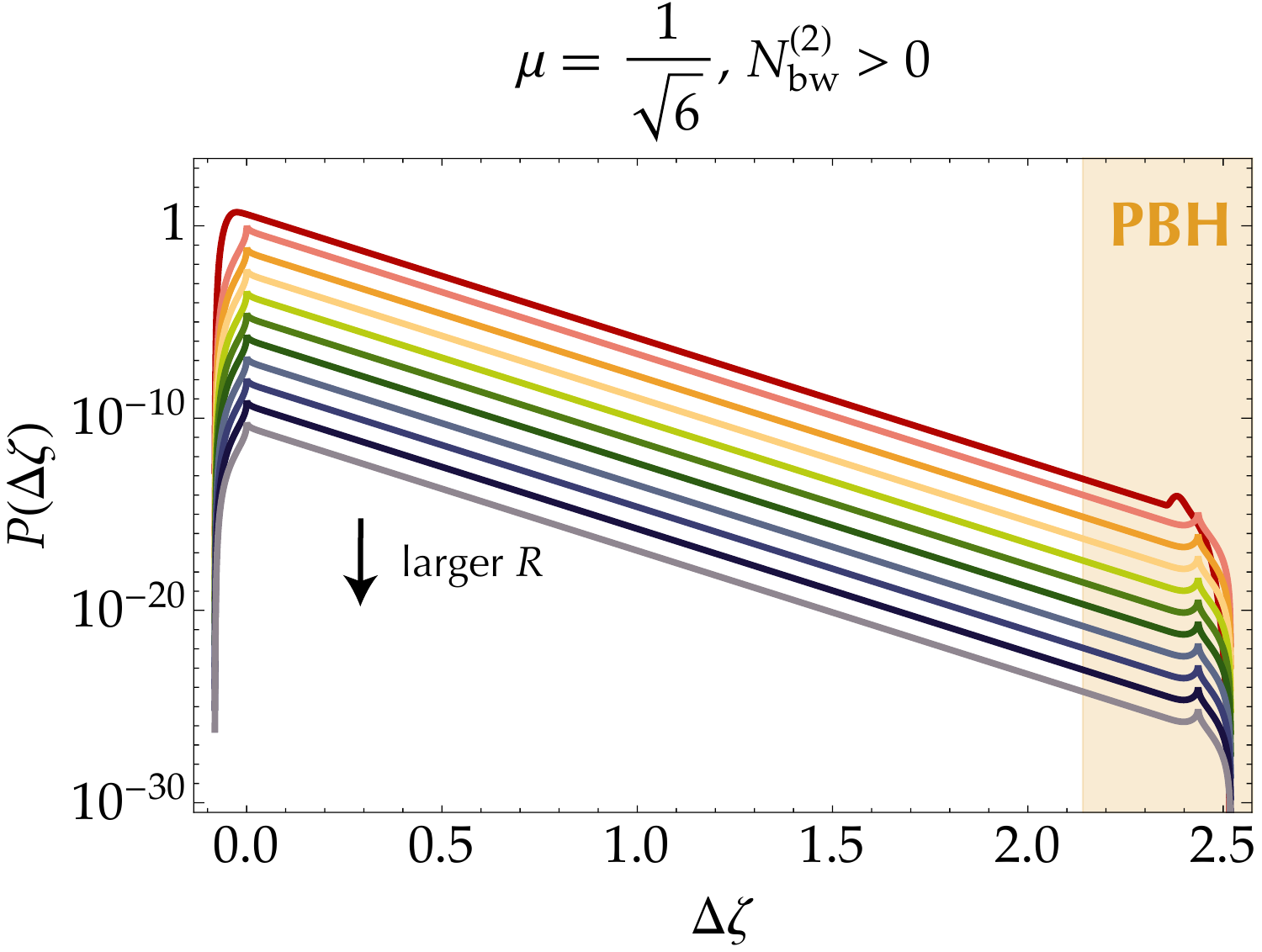}
        \end{minipage}
    \end{tabular}
    \caption{PDF of the coarse-shelled curvature perturbation $\Delta\zeta$ with $\mu=1/\sqrt{6}$ and a few values of the coarse-graining scale $R$ (corresponding to $N_\bw(R)$ in the range $[-0.870,1.57]$ with the equal interval $\Delta N_\bw(R)=0.244$ for the left panel, and the range $[1.57,3.33]$ with $\Delta N_\bw(R)=0.177$ in the right panel). The left panel stands for $N_\bw^{(2)}<0$, \ie it corresponds to the case where $R$ emerges in the second classical slope, while the right panel, where $N_\bw^{(2)}>0$, describes the situation where it does not. The orange shaded region displays the PBH formation criterion~\eqref{eq:Delta:zeta:threshold}. }
    \label{fig: PDz PBH}
\end{figure}
In \Fig{fig: PDz PBH}, we show examples of $P(\Delta\zeta)$ for $N_\bw^{(2)}<0$ (left panel) and $N_\bw^{(2)}>0$ (right panel) with $\mu=1/\sqrt{6}$, varying the coarse-graining scale $N_\bw(R)$. These two regimes feature different behaviours. When $N_\bw^{(2)}<0$ (\ie when $R_2$ emerges in the second classical slope, but not $R_1$), the exponential tail is simply extended up to $\Delta\zeta=N_\bw^{(1)}=N_\bw(R)+\ln\alpha+\ln(1+\beta)$ as the coarse-graining scale $R$ gets larger, while the bulk of the distribution is almost independent of $R$. When $N_\bw^{(2)}>0$ (\ie when neither $R_1$ nor $R_2$ emerge in the second classical slope), the overall amplitude decreases with $R$ [at the rate indicated by \Eq{eq:P2:appr}], while the overall shape remains roughly invariant. The orange shaded region corresponds to the PBH formation criterion, which we further discuss below.
\subsection{Mass function of primordial black holes}
\label{sec:massFraction}
Let us finally investigate the PBH mass function in this simple toy model. Amongst the several approaches that are commonly employed to compute this object, we adopt i) the compaction function in a  radiation-dominated universe ($w=1/3$) for the PBH formation criterion, ii) the critical behaviour for the resultant PBH mass, and iii) the (extended) Press--Schechter approach for the PBH formation probability. Other procedures can be followed, but as stressed above, our goal is to provide an illustration of the formalism introduced in this work, rather than thoroughly studying the formation of PBHs in a realistic model.

The compaction function $\calC(r)$ was introduced around \Eq{eq: compaction function}. Several analytical and numerical works~~\cite{Shibata:1999zs, Harada:2015yda, Musco:2018rwt} suggest that, for a given overdense region, the radius $r_\um$ that maximises the compaction function should be taken as the appropriate coarse-graining scale, and that a PBH forms when the maximum $\calC(r_\um)$ exceeds some almost-universal threshold $\calC_\uth$ (see, \eg, \Refs{Atal:2019erb,Escriva:2019phb} for approaches beyond the mere compaction function). Though the precise value of this threshold has been widely discussed in the literature, in this work we adopt the simple estimate $\calC_\uth\sim w=1/3$ first proposed by Carr in \Refa{Carr:1975qj}, since the other approximations we have performed (namely defining the compaction function with a Gaussian rather than top-hat real-space window function, and using the coarse-shelled curvature perturbation as a proxy) do not allow us to go beyond simple estimates. Through the relation between the compaction function and the coarse-shelled curvature perturbation obtained in \Sec{sec: delta as Dzeta},
\bae{
\label{eq:Deltazeta:C}
    \calC\approx\frac{2}{3}\bqty{1-\pqty{1-\frac{\gamma}{3}\Delta\zeta}^2},
}
the threshold value in terms of $\Delta\zeta$ reads\footnote{Note that, according to \Eq{eq:Deltazeta:C}, $\calC\geq \calC_\uth$ a priori also leads to an upper bound on $\Delta\zeta$, which is however irrelevant (see the discussion around Fig.~3 of \Refa{Kopp:2010sh}).
\label{footnote:Deltazeta_max}}
\bae{
\label{eq:Delta:zeta:threshold}
    \Delta\zeta_\uth\approx\frac{6-3\sqrt{4-6\calC_\uth}}{2\gamma}\simeq 2.14,
}
where $\gamma$ is given by \Eq{eq: alpha beta gamma}. This threshold is shown with the orange regions in \Fig{fig: PDz PBH}.

The resultant PBH mass follows the so-called critical scaling behaviour
\bae{\label{eq: critical behavior}
    M=\kappa M_H(r_\um)\bqty{\calC(r_\um)-\calC_\uth}^p,
}
with the universal scaling index $p=0.36$~\cite{PhysRevLett.70.9,Evans:1994pj,Koike:1995jm,Niemeyer:1997mt,Niemeyer:1999ak,Hawke:2002rf,Musco:2008hv}, and where $\kappa$ is an $\calO(1)$ coefficient that weakly depends on the profile of the overdensity (see, \eg, \Refa{Escriva:2019nsa}). We simply adopt $\kappa\simeq1$ hereafter. The horizon mass $M_H$ at the horizon re-entry of the coarse-graining scale $r_\um$ can be expressed as~(see, \eg, \Refa{Tada:2019amh})\footnote{Strictly speaking, the horizon re-entry is not set by the comoving radius (\ie, $r_\um=1/aH$) but rather by the areal radius $a\ee^{\zeta(r_\um)}r_\um$ (\ie, $a\ee^{\zeta(r_\um)}r_\um=1/H)$~\cite{Yoo:2018kvb,Yoo:2019pma,Yoo:2020dkz,Kitajima:2021fpq}, since the local scale factor is modified by the curvature perturbation. We neglect this effect in this paper for simplicity.}
\bae{
\label{eq:MH(rm)}
    M_H(r_\um)\simeq10^{20}\pqty{\frac{g_*}{106.75}}^{-1/6}
    {\pqty{\frac{r_\um}{6.41\times10^{-14}\,\mathrm{Mpc}}}^2}\,\mathrm{g},
}
where the scale factor is normalised so that its current value is unity, and where we neglect the difference between the effective number of degrees of freedom $g_*$ defined in terms of the energy density and $g_{*s}$, defined in terms of the entropy density. We also uniformly assume $g_*\simeq106.75$ in the mass range of interest, \ie, around $M\sim10^{20}\,\mathrm{g}$ for the numerical application of \Fig{fig: fPBH}.

In a fully non-Gaussian setup, it is difficult to characterise the statistics of $r_\um$ and we therefore proceed as follows. We first fix the coarse-graining scale $r(N_\bw)=R(N_\bw)/a_\uf$, and consider the probability $P(M\mid N_\bw)\dd{\ln M}$ to find PBHs with masses in the range $[M,M\ee^{\dd{\ln M}}]$. Assuming the critical behaviour~\eqref{eq: critical behavior}, in this extended Press--Schechter approach, it is related to the PDF $P(\Delta\zeta\mid N_\bw)$ via
\bae{
    P(M\mid N_\bw)\dd{\ln M}=P(\Delta\zeta\mid N_\bw)\dd{\Delta\zeta}=\frac{9}{4\gamma p}\frac{\calC(\Delta\zeta)-\calC_\uth}{1-\frac{\gamma}{3}\Delta\zeta}P(\Delta\zeta\mid N_\bw)\dd{\ln M}\, .
}
Since one PBH is formed with this probability within each coarse-grained patch of comoving volume $\frac{4\pi}{3}r^3(N_\bw)$, the comoving number density of PBHs in each mass bin at horizon re-entry is given by
\bae{
    n_\PBH(M\mid N_\bw)\dd{\ln M}=
    \frac{27\bqty{\calC(\Delta\zeta)-\calC_\uth}}{16\pi\gamma pr^3(N_\bw)\pqty{1-\frac{\gamma}{3}\Delta\zeta}}P(\Delta\zeta\mid N_\bw)\dd{\ln M}.
}
Neglecting evaporation, accretion and merging, the comoving number density is conserved until today, hence the current ratio between the PBH energy density and the one of dark matter can be written as
\bae{\label{eq: fPBH Nbw}
    &f_\PBH(M\mid N_\bw)\dd{\ln M}=\frac{Mn_\PBH(M\mid N_\bw)}{3\Mpl^2H_0^2\Omega_\DM}\dd{\ln M} \nonumber \\
    &\quad \simeq
    {\frac{\bqty{\calC(\Delta\zeta)-\calC_\uth}^{p+1}}{1-\frac{\gamma}{3}\Delta\zeta}\ee^{-N_\bw}}
    \pqty{\frac{\Omega_\DM h^2}{0.12}}^{-1}
    {\bqty{\frac{r(N_\bw=0)}{6.41\times10^{-14}\,\mathrm{Mpc}}}^{-1}}\bqty{\frac{P(\Delta\zeta\mid N_\bw)}{
    {4.8\times10^{-17}}}}\dd{\ln M},
}
where we adopt the observed value for the current dark matter density $\Omega_\DM h^2\simeq0.12$~\cite{Planck:2018vyg} and note that $r(N_\bw)=r(N_\bw=0)\ee^{N_\bw}$. The normalisation $r(N_\bw=0)$, which determines the typical PBH mass, corresponds to the scale that crosses out the Hubble radius at the onset of the second classical phase and can thus be arbitrarily chosen by tuning the number of classical \efolds spent after the quantum well. 

\bfe{width=0.7\hsize}{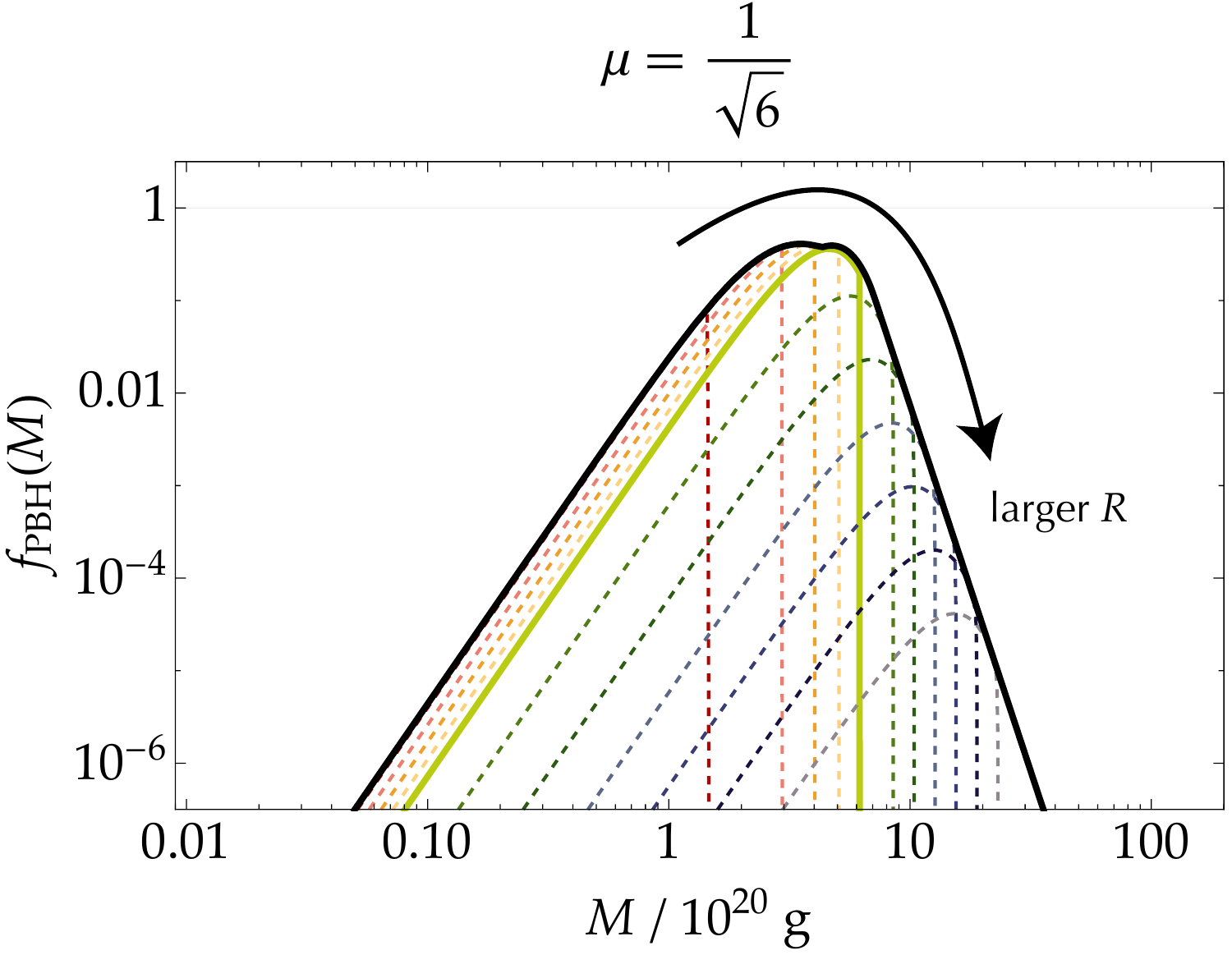}{The PBH mass function with a fixed coarse-graining scale $R$, $f_\PBH[M\mid N_\bw(R)]$ given in \Eq{eq: fPBH Nbw} (coloured lines; $R$ increases from left to right as $r(N_\bw)=(2.51$, $2.64$, $2.77$, $2.91$, $3.06$, $3.38$, $3.74$, $4.13$, $4.57$, $5.05$, $5.58)\times10^{-13}\,\mathrm{Mpc}$), and the full mass function $f_\PBH(M)$ as estimated in \Eq{eq: true fPBH} (black solid line). 
The peaky contributions from $P_1(\Delta\zeta)$ and $\tilde{P}_1(\Delta\zeta)$ are omitted in this plot for simplicity, since they are mere artefacts of having neglected stochastic diffusion in the classical parts of the potential.
We have set  $\mu=1/\sqrt{6}$ and tuned the second classical slope so that $r(N_\bw=0)=6.41\times10^{-14}\,\mathrm{Mpc}$. The light-green solid line corresponds to $N_\bw^{(2)}=0$, \ie to the case where $R^{(2)}$ emerges at the onset of the second classical phase.}{fig: fPBH}
The mass function is displayed in \Fig{fig: fPBH} for $\mu=1/\sqrt{6}$ and for a few values of $N_\bw$, \ie for a few values of $R$, where we have set $r(N_\bw=0)=6.41\times10^{-14}\,\mathrm{Mpc}$. 
In this figure, for simplicity, we did not include the contributions $P_1(\Delta\zeta)$ and $\tilde{P}_1(\Delta\zeta)$, since they give rise to artificial peaks in the mass function which result from having neglected stochastic diffusion in the classical slope and which would be smoothed away otherwise.
The value of $R$ such that $N_\bw^{(2)}=0$ (\ie such that $R^{(2)}$ emerges at the onset of the second classical slope) is shown by the light-green solid line. The dotted curves ranging from dark red to yellow stand for $N_\bw^{(2)}<0$ (\ie $R^{(2)}$ emerges in the second classical slope) and correspond to the situation represented in the left panel of \Fig{fig: PDz PBH}. As $R$ increases, the PDF of $\Delta\zeta$ is roughly invariant, apart from the location of the hard cutoff that is driven to larger values. This is why, in \Fig{fig: fPBH}, the overall shape of the mass distribution is almost constant in this regime, with a mere shift in mass. In contrast, the dotted curves ranging from green to dark blue and grey stand for $N_\bw^{(2)}>0$ (\ie neither $R^{(1)}$ nor $R^{(2)}$ emerge in the second classical slope) and correspond to the situation represented in the left panel of \Fig{fig: PDz PBH}. As $R$ increases, the PDF of $\Delta\zeta$ gets suppressed [roughly by a factor $\ee^{-\pi^2 N_\bw(R)/(4\mu^2)}$, see \Eq{eq:P2:appr}], which leads to the suppression observed in \Fig{fig: fPBH} (together with the mass sift). 
One notices that there is a small discontinuity at $N_\bw^{(2)}=0$ just before the light-green line. This is caused by the discontinuous statistics of $x_*^{(2)}$: it is fixed in the second classical slope for $N_\bw^{(2)}<0$, while it can be broadly distributed in the quantum well for $N_\bw^{(2)}>0$. This artefact would be also smoothed away by properly accounting for stochastic diffusion in the classical parts of the potential.

Let us also note that while the low-mass tail is caused by the critical behaviour~\eqref{eq: critical behavior} (see, \eg, \Refa{Kitajima:2021fpq}), the high-mass tail originates from the ``contamination" effect discussed around \Fig{fig: Pwell zR}: even at scales much larger than those typically emerging in the quantum well, the heavy tails generated in the quantum well leave a non-trivial imprint. This implies that PBHs may form with masses much larger than those naively expected in the quantum well, an interesting result indeed.

Finally, the full mass distribution is formed from $f_\PBH[M\mid N_\bw(r_\um)]$ by accounting for the actual value of $r_\um$ in each patch. As mentioned above, the characterisation of $r_\um$ in a fully non-Gaussian setup is a non-trivial task, and we leave it for future work. Here we simply note that, in practice, most PBHs of a given mass $M$ arise from patches with the same size $r_\um$. In other words, if one studies the $f_\PBH[M\mid N_\bw(r_\um)]$ as a function of $r_\um$ when $M$ is fixed, one realises that this function features a very sharp maximum at a certain value $r_\um^\umax(M)$. If one assumes that all PBHs of mass $M$ start from patches of that size (\ie if one neglects the contribution from patches of other sizes), one can approximate $f_\PBH(M)\simeq f_\PBH\{M\mid N_\bw[r_\um^\umax(M)]\}$, \ie,\footnote{The condition appearing in \Eq{eq: true fPBH} guarantees that $N_\bw^{(1)}=N_\bw+\ln\alpha+\ln(1+\beta)$ is positive, since otherwise the two scales $R^{(1)}$ and $R^{(2)}$ emerge in the second classical slope and the coarse-shelled curvature perturbation vanishes (given that we have neglected stochastic diffusion in that part).}
\bae{\label{eq: true fPBH}
    f_\PBH(M)=\max\bqty{f_\PBH(M\mid N_\bw)\mid N_\bw>-\ln\alpha-\ln(1+\beta)}.
}
This implies that $f_\PBH(M)$ can be approximated by the envelope curve of the functions $f_\PBH(M\mid N_\bw)$ displayed in \Fig{fig: fPBH}, where $f_\PBH(M)$ is shown with the solid black line.  

The mass distribution is moderately broad as it extends over a couple of decades, around a certain maximum. The location of the maximum depends on the normalisation $r(N_\bw=0)$, which can be freely adjusted by tuning the number of classical \efolds realised after the quantum well, as mentioned above. For masses smaller than the maximum, the slope of the mass function directly reflects the critical scaling behaviour~\eqref{eq: critical behavior}, since all masses located on the lower tail roughly come from the same scale $r_\um$ (this scale corresponds to the dark-red dotted line in \Fig{fig: fPBH} to which the black solid line indeed asymptotes, and it also roughly corresponds to the dark-violet line in the left panel of \Fig{fig: PDz PBH}, \ie, to the smallest value of $r_\um$ such that $P(\Delta\zeta)$ intersect the PBH formation region in that figure). For masses larger than the maximum, \ie for the upper tail in \Fig{fig: fPBH}, the decay rate is more directly driven by the ``contamination effect'' mentioned above, \ie, it is driven by stochastic effects. 

Finally, we stress that this represents the first derivation of a PBH mass function in a full stochastic analysis. PBH abundances were estimated before, but the mass distribution itself requires to properly account for the relationship between masses, length scales, and the field configuration when those length scales emerge, which is not possible without the framework introduced in this paper.
\section{Conclusions}
\label{sec:conclusion}
In this work, we derived a generic framework to compute the one-point statistics of cosmological perturbations when coarse-grained at an arbitrary scale, in the presence of quantum diffusion. This bridges the final gap between the stochastic-inflation formalism and the calculation of the mass distribution of astrophysical objects such as primordial black holes. 

In practice, this was done by relating the curvature perturbation $\zeta_R$, coarse-grained at a physical distance $R$, to the integrated amount of expansion realised until that scale $R$ crosses out the Hubble radius during inflation. Using the first-passage time techniques of the stochastic-$\delta N$ formalism, we derived the relevant formulae to compute the probability density function of this quantity. We also pointed out that other relevant cosmological fields, such as the comoving density contrast or the compaction function, can be approximated by the difference in the curvature perturbation when coarse-grained at two different scales $R_1$ and $R_2$. This allowed us to generalise our results to the calculation of the one-point statistics of such cosmological fields, which are better suited to compute the abundance of extreme objects such as primordial black holes.

To illustrate how the formalism can be employed in practice, we then applied those formulae to a toy model where inflation is driven by a single scalar field, the potential of which contains an exactly flat region. In this ``quantum-well'' model, we found that cosmological perturbations feature a heavy, highly non-Gaussian tail, confirming similar observations previously made at the level of the first-passage time distributions. More precisely, we found that the tail of the PDF of the curvature perturbation is exponential with an additional cubic suppression, while the tail of the PDF of the density contrast and of the compaction function is merely exponential. 

We then applied those results to the derivation of the mass fraction of primordial black holes in this model. We found that, while the low-mass end of the distribution directly reflects the critical scaling behaviour (namely the relationship between the amplitude of the initial overdensity and the mass of the resultant black hole), the large-mass end is mostly driven by stochastic diffusion. In particular, it reveals the presence of a ``contamination effect'' already unveiled at the level of the power spectrum in \Refa{Ando:2020fjm}: even at scales much larger than those typically emerging in the quantum well, the heavy tails imply that PBHs may form, hence with masses much larger than those naively expected in this model.

\bigskip

The main result of this paper lies in the construction of a generic formalism to derive the one-point statistics of cosmological fields when coarse-grained at an arbitrary scale. Those statistics serve as the starting point of the calculation of various cosmological observables, such as the mass function of astrophysical compact objects. This is why a natural prospect of the present work is the application of this formalism to various setups. The ``quantum-well'' example we have analysed here is only a toy model, and it would be interesting to see whether and how the conclusions derived in this setup generalise to more realistic scenarios, possibly with multiple-field effects, phases of ultra-slow roll, etc. It would also be interesting to incorporate extensions of the stochastic-inflation equations beyond the leading order in sub-Hubble interactions (see, e.g., \Refs{Tokuda:2017fdh,Tokuda:2018eqs,Gorbenko:2019rza,Mirbabayi:2020vyt,Cohen:2021fzf}) into our formalism.

It is important to notice that our ability to express the one-point statistics of cosmological fields in terms of first-passage time distributions  (and hence the possibility to derive analytical results) only relies on a Markovian assumption. Fortunately, in the slow-roll regime, stochastic inflation is described by Markovian processes. The reason is that the statistical properties of the noise are determined by the few \efolds surrounding the Hubble crossing time of the Fourier comoving modes that contribute to the noise. If the local background geometry does not evolve much during this period, the properties of the noise only depend on the local configuration of the background at the time it emerges, hence the process is Markovian. In the presence of sudden and transient departures from the slow-roll attractor, this approximation may break down, but even in that case  non-Markovian effects have been found to be subleading in \Refa{Figueroa:2021zah}. The first-passage time techniques of the stochastic-$\delta N$ formalism seem therefore to be applicable to a broad range of setups. Note that if one wanted to account for non-Markovian effects, one could still do it with the generic equation~\eqref{eq:zetaR:vol:averages}, and making use of stochastic lattice simulations.

It is finally worth pointing out that, once the one-point statistics of the relevant cosmological fields is computed, the estimation of the mass distribution of primordial black holes can be performed with different levels of refinement. In the example treated in this work, we used the simple Press--Schechter procedure as an illustration, but it would be interesting to see how the heavy tails we have encountered are processed by more advanced methods (such as peak theory --- following the lines of \Refs{Yoo:2018kvb,Yoo:2019pma,Yoo:2020dkz,Kitajima:2021fpq}, or even in numerical simulations).

\acknowledgments

We are grateful to Hooshyar Assadullahi, Andrew Gow, Joseph Jackson, Kazuya Koyama, Junsei Tokuda, and David Wands for helpful discussions and Baptiste Blachier for pointing out typos.
Y.T. is supported by JSPS KAKENHI Grants No. JP19K14707 and No. JP21K13918.

\appendix

\section{Formulae in the quantum-well model}
\label{app:quantum:well}\allowdisplaybreaks
In this appendix, we present the detailed calculation of the formulae presented in \Sec{sec:QuantumWell} for the quantum-well toy model. Our starting point is the solution of the Fokker--Planck and first-passage-time problems respectively given in \Eqs{eq: P qwell} and~\eqref{eq: Pfpt qwell}.

The backward probability $P_\bw$ can be calculated by plugging these results into \Eq{eq: Pbw}, which reduces to (see Appendix A of \Refa{Ando:2020fjm} for a detailed derivation)
\bae{\label{eq: Pbw qwell}
    &P_\bw\left[x\mid N_\bw(R)\right] \nonumber \\ &
    \simeq\bce{
        \dps N_\cl^\prime(x)
        P_\FPT^\uwell\left[N_\ubw(R)-N_\ucl(x)\mid 1,\mu\right]
        \theta \left[{N_\bw(R)-N_\cl(x)}\right]\theta(x_0-x) &\ \text{if}\ x> 1 \\
        \dps
        \mu^2 x P_\FPT^\uwell\left[N_\ubw(R)\mid x,\mu\right] &\ \text{if}\ x\leq1
    } ,
}
where we recall that $P_\FPT^\uwell$ is given in \Eq{eq: Pwell FPT}. 
In this expression, we have approximated the denominator of \Eq{eq: Pbw} by unity, which amounts to assuming $N_\bw(R)\ll\braket{\calN(x_0)}$, and which is valid when the initial condition for our universe $x_0$ is chosen sufficiently high up in the potential. We have also used a simplified notation $N_\cl(x)\coloneqq N_\cl(1;x)$.

The integral that appears in the expression~\eqref{eq: PzetaR exact} for $P(\zeta_R)$ can be split into two parts as
\bae{
    P(\zeta_R)&=\int_0^1\dd{x_*}P_\bw[x_*\mid N_\bw(R)]\,P_\FPT[\braket{\calN(x_0)}-\braket{\calN(x_*)}+\zeta_R\mid x_0\to x_*] \nonumber \\
    &\quad+\int_1^{x_0}\dd{x_*}P_\bw[x_*\mid N_\bw(R)]\,P_\FPT[\braket{\calN(x_0)}-\braket{\calN(x_*)}+\zeta_R\mid x_0\to x_*].
}
Here, the first-passage-time PDF $P_\FPT(\calN\mid x_0\to x_*)$, with the shifted end-point $x_*>0$, is related to the ``original'' one where the end-point is $x=0$ by
\bae{
    P_\FPT(\calN\mid x_0\to x_*)=\bce{
        \dps
        \delta[\calN-N_\cl(x_*;x_0)] &\qquad\text{if}\quad x_*>1 \\
        \dps
        P_\FPT[\calN-N_\cl(1;x_0)\mid1,(1-x_*)\mu]&\qquad\text{if}\quad x_*\leq1
    },
}
with $P_\FPT$ given in \Eq{eq: Pfpt qwell}. Moreover, the mean number of \efolds~$\braket{\calN(x)}$ can be obtained by solving the first partial differential equation in \Eq{eq: recursive PDE} (\ie~the one with $n=1$), and one finds
\bae{\label{eq: mean N}
    \braket{\calN(x)}=\bce{
        \dps
        \frac{\mu^2}{2}+N_\cl(1;x)&\qquad\text{if}\quad x>1 \\[5pt]
        \dps
        \frac{\mu^2}{2}\bqty{1-(1-x)^2}&\qquad\text{if}\quad x\leq 1
    }.
}
Combining the above results, one obtains \Eq{eq: PzetaR qwell} for $P(\zeta_R)$.

Let us now discuss the coarse-shelled curvature perturbation, which as explained in \Sec{sec:delta:C} provides a proxy for the density contrast and for the compaction function. Plugging the above formulae into \Eq{eq: joint Pbw}, one obtains for the joint backward probability
\bae{\label{eq: Pbw 0<x1,x2<1}
    &P_\bw\left(x_*^{(1)},x_*^{(2)}\relmiddle{|} N_\bw^{(1)},N_\bw^{(2)}\right)=-\frac{\pi x_*^{(1)}}{4}\vartheta_2^\prime\pqty{\frac{\pi}{2}x_*^{(2)},\ee^{-\frac{\pi^2N_\bw^{(2)}}{\mu^2}}} \nonumber \\
    &\quad\times\Bqty{\vartheta_2\bqty{\frac{\pi}{2}(x_*^{(1)}-x_*^{(2)}),\ee^{-\frac{\pi^2\ln(1+\beta)}{\mu^2}}}-\vartheta_2\bqty{-\frac{\pi}{2}(x_*^{(1)}+x_*^{(2)}),\ee^{-\frac{\pi^2\ln(1+\beta)}{\mu^2}}}}
}
for $0<x_*^{(1)},x_*^{(2)}<1$, 
\bae{\label{eq: Pbw 0<x2<1<x1}
    &P_\bw\left(x_*^{(1)},x_*^{(2)}\relmiddle{|} N_\bw^{(1)},N_\bw^{(2)}\right)=-\frac{\pi N_\cl^\prime\pqty{x_*^{(1)}}}{4\mu^2}\vartheta_2^\prime\pqty{\frac{\pi}{2}x_*^{(2)},\ee^{-\frac{\pi^2N_\bw^{(2)}}{\mu^2}}} \nonumber \\
    &\times\Bqty{\vartheta_2\bqty{-\frac{\pi}{2}(x_*^{(2)}-1),\ee^{-\frac{\pi^2\pqty{\ln(1+\beta)-N_\cl\pqty{x_*^{(1)}}}}{\mu^2}}}-\vartheta_2\bqty{-\frac{\pi}{2}(x_*^{(2)}+1),\ee^{-\frac{\pi^2\pqty{\ln(1+\beta)-N_\cl\pqty{x_*^{(1)}}}}{\mu^2}}}} \nonumber \\
    &\times\theta\bqty{\ln(1+\beta)-N_\cl\pqty{x_*^{(1)}}}
}
for $0<x_*^{(2)}<1<x_*^{(1)}<x_0$,
\bae{\label{eq: Pbw 1<x2<x1}
    &P_\bw\left(x_*^{(1)},x_*^{(2)}\relmiddle{|} N_\bw^{(1)},N_\bw^{(2)}\right)=-\frac{\pi N_\cl^\prime\pqty{x_*^{(1)}}}{2\mu^2}\vartheta_2^\prime\Bqty{\frac{\pi}{2},\ee^{-\frac{\pi^2\bqty{N_\bw^{(2)}-N_\cl\pqty{x_*^{(2)}}}}{\mu^2}}} \nonumber \\
    &\quad\times\delta\Bqty{x_*^{(2)}-x_\cl\bqty{\ln(1+\beta);x_*^{(1)}}}\theta\bqty{N_\bw^{(1)}-N_\cl\pqty{x_*^{(1)}}}\theta\bqty{N_\bw^{(2)}-N_\cl\pqty{x_*^{(2)}}}
}
for $1<x_*^{(2)}<x_*^{(1)}<x_0$, and $P_\bw\left(x_*^{(1)},x_*^{(2)}\relmiddle{|}N_\bw^{(1)},N_\bw^{(2)}\right)=0$ otherwise.

Making use of \Eq{eq: mean N} for the mean number of \efolds, the integration over $x_*^{(2)}$ that appears in \Eq{eq: PDeltaz} can be performed by means of the Dirac distribution, leading to
\bae{
    &P(\Delta\zeta) \nonumber \\
    &=\bigintss\dd{x_*^{(1)}}\eval{\frac{P_\bw\left(x_*^{(1)},x_*^{(2)}\relmiddle{|}N_\bw^{(1)},N_\bw^{(2)}\right)}{\mu^2\pqty{1-x_*^{(2)}}}\theta\bqty{0<f\pqty{x_*^{(1)},\Delta\zeta}<\frac{\mu^2}{2}}}_{x_*^{(2)}=1-\sqrt{1-\frac{2f\pqty{x_*^{(1)},\Delta\zeta}}{\mu^2}}} \nonumber \\
    &\quad+\bigintss\dd{x_*^{(1)}}\eval{\frac{P_\bw\left(x_*^{(1)},x_*^{(2)}\relmiddle{|}N_\bw^{(1)},N_\bw^{(2)}\right)}{N_\cl^\prime\pqty{x_*^{(2)}}}\theta\bqty{f\pqty{x_*^{(1)},\Delta\zeta}>\frac{\mu^2}{2}}}_{x_*^{(2)}=x_\cl\bqty{f\pqty{x_*^{(1)},\Delta\zeta}-\frac{\mu^2}{2}}},
}
where we have defined
\bae{
    f\pqty{x_*^{(1)},\Delta\zeta}\coloneqq\Braket{\calN\pqty{x_*^{(1)}}}-\ln(1+\beta)+\Delta\zeta,
}
and used a simplified notation $x_\cl(N)$ as the inverse function of $N_\cl(x)=N_\cl(1;x)$.
In order to make use of the explicit expressions of $P_\bw$, \ie~\Eqs{eq: Pbw 0<x1,x2<1}--\eqref{eq: Pbw 1<x2<x1}, we then split the integration over $x_*^{(1)}$ as
\bae{
    &P(\Delta\zeta) \nonumber \\
    &=\bigintss_0^1\dd{x_*^{(1)}}\eval{\frac{P_\bw\left(x_*^{(1)},x_*^{(2)}\relmiddle{|}N_\bw^{(1)},N_\bw^{(2)}\right)}{\mu^2\pqty{1-x_*^{(2)}}}\theta\bqty{0<f\pqty{x_*^{(1)},\Delta\zeta}<\frac{\mu^2}{2}}}_{x_*^{(2)}=1-\sqrt{1-\frac{2f\pqty{x_*^{(1)},\Delta\zeta}}{\mu^2}}} \nonumber \\
    &\quad+\bigintss_1^{x_0}\dd{x_*^{(1)}}\eval{\frac{P_\bw\left(x_*^{(1)},x_*^{(2)}\relmiddle{|}N_\bw^{(1)},N_\bw^{(2)}\right)}{\mu^2\pqty{1-x_*^{(2)}}}\theta\bqty{0<f\pqty{x_*^{(1)},\Delta\zeta}<\frac{\mu^2}{2}}}_{x_*^{(2)}=1-\sqrt{1-\frac{2f\pqty{x_*^{(1)},\Delta\zeta}}{\mu^2}}} \nonumber \\
    &\quad+\bigintss_1^{x_0}\dd{x_*^{(1)}}\eval{\frac{P_\bw\left(x_*^{(1)},x_*^{(2)}\relmiddle{|}N_\bw^{(1)},N_\bw^{(2)}\right)}{N_\cl^\prime\pqty{x_*^{(2)}}}\theta\bqty{f\pqty{x_*^{(1)},\Delta\zeta}>\frac{\mu^2}{2}}}_{x_*^{(2)}=x_\cl\bqty{f\pqty{x_*^{(1)},\Delta\zeta}-\frac{\mu^2}{2}}}.
}
Recalling that $N_\bw^{(1)}=N_\bw^{(2)}+\ln(1+\beta)=N_\bw(R)+\ln\alpha+\ln(1+\beta)$ and performing the change of integration variable $x_*^{(1)}\to N^{(1)}=N_\cl\pqty{x_*^{(1)}}$ in the last two terms, one finds 
\bae{
    &P(\Delta\zeta) \nonumber \\
    &=-\frac{\pi}{4\mu^2}\theta\bqty{\ln(1+\beta)-\frac{\mu^2}{2}<\Delta\zeta<\ln(1+\beta)+\frac{\mu^2}{2}} \nonumber \\
    &\qquad\times\bigintss_{\max\Bqty{0,1-\sqrt{1+\frac{2}{\mu^2}\bqty{\Delta\zeta-\ln(1+\beta)}}}}^{1-\sqrt{\max\Bqty{0,\frac{2}{\mu^2}\bqty{\Delta\zeta-\ln(1+\beta)}}}}\dd{x_*^{(1)}}\frac{x_*^{(1)}}{1-x_\well^{(2)}\left(x_*^{(1)}\right)}\vartheta_2^\prime\Bqty{\frac{\pi}{2}x_\well^{(2)}\left(x_*^{(1)}\right),\ee^{-\frac{\pi^2\bqty{N_\bw(R)+\ln\alpha}}{\mu^2}}} \nonumber \\
    &\qquad\quad\times\pqty{\vartheta_2\Bqty{\frac{\pi}{2}\bqty{x_*^{(1)}-x_\well^{(2)}\left(x_*^{(1)}\right)},\ee^{-\frac{\pi^2\ln(1+\beta)}{\mu^2}}}-\vartheta_2\Bqty{-\frac{\pi}{2}\bqty{x_*^{(1)}+x_\well^{(2)}\left(x_*^{(1)}\right)},\ee^{-\frac{\pi^2\ln(1+\beta)}{\mu^2}}}} \nonumber \\
    &\quad-\frac{\pi}{4\mu^4}\theta\bqty{-\frac{\mu^2}{2}<\Delta\zeta<\ln(1+\beta)} \nonumber \\
    &\qquad\times\bigintss_{\max\bqty{0,\ln(1+\beta)-\Delta\zeta-\frac{\mu^2}{2}}}^{\ln(1+\beta)-\max(0,\Delta\zeta)}\dd{N^{(1)}}\frac{\vartheta_2^\prime\Bqty{\frac{\pi}{2}x_\cl^{(2)}(N^{(1)}),\ee^{-\frac{\pi^2\bqty{N_\bw(R)+\ln\alpha}}{\mu^2}}}}{1-x_\cl^{(2)}(N^{(1)})} \nonumber \\
    &\qquad\quad\times\Bigg(\vartheta_2\Bqty{-\frac{\pi}{2}\bqty{x_\cl^{(2)}(N^{(1)})-1},\ee^{-\frac{\pi^2\bqty{\ln(1+\beta)-N^{(1)}}}{\mu^2}}} \nonumber \\
    &\qquad\qquad\qquad-\vartheta_2\Bqty{-\frac{\pi}{2}\bqty{x_\cl^{(2)}(N^{(1)})+1},\ee^{-\frac{\pi^2\bqty{\ln(1+\beta)-N^{(1)}}}{\mu^2}}}\Bigg) \nonumber \\
    &\quad-\frac{\pi}{2\mu^2}\delta(\Delta\zeta)\int_{\ln(1+\beta)}^{N_\bw(R)+\ln\alpha+\ln(1+\beta)}\dd{N^{(1)}}\vartheta_2^\prime\Bqty{\frac{\pi}{2},\ee^{-\frac{\pi^2\bqty{N_\bw(R)+\ln\alpha+\ln(1+\beta)-N^{(1)}}}{\mu^2}}},
}
where 
\beae{
    &x_\well^{(2)}\pqty{x_*^{(1)}}\coloneqq1-\sqrt{\pqty{1-x_*^{(1)}}^2-\frac{2}{\mu^2}\bqty{\Delta\zeta-\ln(1+\beta)}}, \\
    &x_\cl^{(2)}\pqty{N^{(1)}}\coloneqq1-\frac{\sqrt{2}}{\mu}\sqrt{\ln(1+\beta)-\Delta\zeta-N^{(1)}}.
}
Note that we explicitly included step functions in the first and the second terms to ensure that the integration ranges are positively oriented. These expressions can be further simplified by making use of the periodic relations $\vartheta_2(z+\pi/2,q)=-\vartheta_1(z,q)$ and $\vartheta_1(z+\pi/2,z)=\vartheta_2(z,q)$, and one finally obtains \Eqs{eq:P:Deltazeta:quantum:well:gen} and~\eqref{eq: P1 P2 P3}. There, we have changed the integration variable of the first term, $P_1(\Delta\zeta)$, according to $x_*^{(1)}\to \tilde{x}_*^{(1)}\coloneqq1-x_*^{(1)}$, and of the third term, $P_3(\Delta\zeta)$, according to $N^{(1)}\to N^{(2)}\coloneqq N^{(1)}-\ln(1+\beta)$.

\bibliographystyle{JHEP}
\bibliography{main}
\end{document}